\documentclass{aa}  

\usepackage{graphicx}
\usepackage{txfonts}
\usepackage{nag}
\usepackage{amsmath, amsfonts, amssymb}
\usepackage{bm}
\usepackage{hyperref}
\usepackage{pdflscape}
\usepackage{placeins}
\usepackage{subcaption}
\usepackage{xspace}
\usepackage{hyphenat}
\usepackage{natbib}

\newcommand{\hi}{\textsc{H$\,$i}\xspace}
\newcommand{\unit}[2]{\ensuremath{\text{#1}^{#2}}\xspace}

\newcommand{\numunit}[2]{\mbox{\ensuremath{#1\,#2}\xspace}}

\newcommand{\mstar}{\ensuremath{\text{M}_\star}\xspace}
\newcommand{\mhi}{\ensuremath{\text{M}_\hi}\xspace}

\newcommand{\kms}{\ensuremath{\text{km}\,\text{s}^{-1}}\xspace}
\newcommand{\msol}{\ensuremath{\text{M}_\odot}\xspace}

\newcommand{\ave}[1]{\ensuremath{\langle #1 \rangle}\xspace}
\newcommand{\fhi}{\ensuremath{f_\hi}\xspace}

\newcommand{\gmr}{\ensuremath{g-r}\xspace}

\newcommand{\appendref}[1]{Appendix~\ref{#1}\xspace}
\newcommand{\iraf}{IRAF\xspace}
\newcommand{\defhi}{\ensuremath{\textsc{Def}_\hi}\xspace}
\graphicspath{{images/}}

\newcommand{\cittmolnar}{Molnar et al. (in prep)\xspace}

\newcommand{\secref}[1]{Sec.~\ref{#1}}
\newcommand{\figref}[1]{Fig.~\ref{#1}}
\newcommand{\tabref}[1]{Table~\ref{#1}}

\begin{document}

   \title{H$\,\textsc{i}$ content in Coma cluster substructure\thanks{Appendices C and D are only available in electronic format at the CDS via anonymous ftp to cdsarc.u-strasbg.fr (130.79.128.5) or via http://cdsweb.u-strasbg.fr/cgi-bin/qcat?J/A+A/}.}
   \titlerunning{H$\,\textsc{i}$ content in Coma cluster substructure}
   \authorrunning{J. Healy et al.}

   \author{J. Healy,
          \inst{1,2}\thanks{contact: healy@astro.rug.nl}
          S-L. Blyth,\inst{2} 
          M.A.W. Verheijen,\inst{1}
          K.M. Hess,\inst{1,3} 
          P. Serra,\inst{4} 
          J.M. van der Hulst,\inst{1,3} 
          T.H. Jarrett,\inst{2} 
          K. Yim\inst{5} 
          \and
          G. I. G. J\'ozsa\inst{6,7,8}
          }

   \institute{Kapteyn Astronomical Institute, University of Groningen, Landleven 12, 9747 AV Groningen, the Netherlands
         \and
         Department of Astronomy, University of Cape Town, Private Bag X3, Rondebosch 7701, South Africa
         \and 
         Netherlands Institute for Radio Astronomy (ASTRON), Oude Hoogeveensedijk 4, 7991 PD Dwingeloo, the Netherlands
         \and
         INAF - Osservatorio Astronomico di Cagliari, Via della Scienza 5, I-09047 Selargius (CA), Italy
         \and
         Korea Astronomy and Space Science, Institute, 776 Daedeok-daero, Yuseong-gu, Daejeon 34055, Korea
         \and
         South African Radio Astronomy Observatory, 2 Fir Street, Black River Park, Observatory, Cape Town, 7925, South Africa 
         \and
        Department of Physics and Electronics, Rhodes University, PO Box 94, Makhanda, 6140, South Africa
         \and
        Argelander-Institut f\"ur Astronomie, Auf dem H\"ugel 71, D-53121 Bonn, Germany\\
             }

   \date{Received June 24, 2020; accepted November 3, 2020}

 
  \abstract
   {Galaxy clusters are some of largest structures in the universe. These very dense environments tend to be home to higher numbers of evolved galaxies that what is found in lower density environments. It is well known that dense environments can influence the evolution of galaxies through the removal of the neutral gas (\hi) reservoirs which fuel star formation. It is unclear which environment has a stronger effect: the local environment (i.e. the substructure within the cluster), or the cluster itself. }
   {Using the new \hi data from the Westerbork Coma Survey, we explore the average \hi content of galaxies across the cluster comparing galaxies that reside in substructure to those that do not.}
   {We apply to the Dressler-Shectman test to our newly compiled redshift catalogue of the Coma cluster to search for substructure. With so few of the Coma galaxies directly detected in \hi, we use the \hi stacking technique to probe average \hi content below what can be directly detected.}
   {Using the Dressler-Shectman test, we find 15 substructures within the footprint of the Westerbork Coma Survey. We compare the average \hi content for galaxies within substructure to those not in substructure. Using the \hi stacking technique, we find that the Coma galaxies (for which are not detected in \hi) are more than 10--50 times more \hi deficient than expected which supports the scenario of an extremely efficient and rapid quenching mechanism. By studying the galaxies that are not directly detected in \hi, we also find Coma to be more \hi deficient than previously thought. }
   {}


   \maketitle
%

\section{Introduction}
\label{sec:coma_intro}
    
    In our current picture of cosmology, $\Lambda$~cold dark matter ($\Lambda$CDM), larger structures (e.g. galaxy clusters) are built up through the hierarchical merging of smaller structures (e.g. galaxy groups or individual galaxies) \citep{Springel2005}. Using dark matter simulations, \citet{McGee2009} showed that 45 per cent of galaxies accreted onto a Coma-sized cluster would have done so in groups. Their results suggest that the group environments play an important role in the evolution of galaxies prior to accretion onto the cluster. Traces of a cluster's accretion history can be observed through substructure in the kinematics of the galaxies in the cluster \citep[e.g.][]{Dressler1988,Hou2012}. Observations have shown that roughly 30 per cent of clusters contain kinematic substructure \citep{Dressler1988,Bird1994,Hou2012}. The observational evidence of substructure is further supported by dark matter simulations which estimate a similar fraction of clusters contain significant substructure \citep{Knebe1999}. \\
 
    Given the prevalence of substructure within clusters, there have been many efforts to identify substructure. Groups of galaxies identified through substructure within the cluster's x-ray emission (e.g. \citealt{Neumann2001, Neumann2003,Parekh2015,Zhang2009}) may be older, more evolved groups that have had their own intracluster medium (ICM) prior to accretion onto the cluster. An alternative possibility is that the x-ray emission associated with such groups might not belong to them but rather to cluster ICM that was disturbed by the fast peri-centre passage of such recently accreted groups. Another identification method is a hierarchical clustering technique, a mathematical algorithm which sorts data into `clusters' based on an affinity parameter. This particular method, introduced by \citet{Serna1996}, assigns galaxies to groups by minimizing the binding energy. However, one of the more common methods of identifying substructure is to look for kinematic deviations ($\delta$ values) between groups of galaxies and the cluster as a whole: this method is known as the Dressler-Shectman (DS) Test \citep{Dressler1988}. Aside from the original study where \citet{Dressler1988} looked for kinematic substructures in the Abell cluster catalogue, the DS test has been used to study the substructure in Abell 2192 and Abell 963 \citep{Jaffe2013}, the Antlia cluster \citep{Hess2015}, and the Coma cluster \citep{Colless1996}. All of the above methods are effective at finding substructure, however each method is sensitive to different density environments within a cluster. \citet{Knebe1999} studied the effectiveness of the DS test using numerical simulations of clusters and also compared the DS test to other methods such as the friends-of-friends (FoF) algorithm. They found that the DS test and FoF algorithm find similar structures, but there is a large scatter. \citet{Knebe1999} attribute the scatter (and also the differences in subgroups for the same cluster in the literature) to velocity projection effects which can artificially increase the $\delta$ values in the DS test.\\
    
    \citet{Neumann2003} identified substructure in the Coma cluster by finding over-densities in the x-ray emission in the cluster. At the core of the cluster, they found an x-ray over-density associated with the two cD galaxies (NGC 4889 and NGC 4874).{ This x-ray over-density is offset from the x-ray centre of the cluster and has been attributed to the halos of the individual galaxies \citep{Andrade-Santos2013,Vikhlinin2001}. Previous kinematic substructure analyses \citep[e.g.][]{Colless1996} have identified separate groups associated with NGC 4889 and NGC 4874. \citet{Colless1996} and \citet{White1993} suggest that the NGC4889 group is merging with the NGC 4874 group which was dislodged from the centre of the cluster potential due to the interaction. This would account for the velocity differences between the two galaxies themselves and also with respect to the cluster velocity.} \\
    
    Studies using the DS test and hierarchical clustering technique have found groups throughout the cluster, but both methods are sensitive to the availability of optical redshifts for a cluster \citep{Adami2005}. This is particularly evident when comparing the first DS test results from \citet{Dressler1988} who found no significant substructure within the Coma cluster (their catalogue contained 100--200 galaxies over a \numunit{2}{\text{deg}^2} area), with later work on the Coma cluster by \citet{Colless1996}, who found that there is significant substructure with a larger redshift catalogue over a similar area of the sky.  \\

    Identifying substructure within galaxy clusters is not only important for understanding the build up of the cluster, the local environments identified by substructure also provide insight into the evolution of galaxies \citep{Neumann2001,Neumann2003,Dressler1988,Adami2005,Hess2015}. \citet{Dressler1980} found that galaxies residing in dense environments tend to be passive with more evolved morphologies, while low density environments have higher fractions of less evolved, actively star-forming galaxies. This observed phenomenon is now referred to as the morphology-density relation. Uncovering the cause of the morphology-density relation is related to understanding the evolutionary pathway from late-type spiral (Sa/Sb/Sc/Sd) galaxies to early-type elliptical (dE/E) galaxies. On the traditional Hubble galaxy morphology ladder, lenticular (S0) galaxies bridge the gap between the ellipticals and the spirals. While S0s are thought to be quenched spirals \citep[see][and references therein]{Coccato2019}, it is not known how or even if S0s transition to elliptical galaxies. In cluster environments where the fraction of S0 galaxies tends to be high, gas removal processes dominate the evolutionary mechanisms \citep[see][for a review]{Aguerri2012}. It is still unclear which environment has a stronger effect in driving morphological evolution -- does the cluster drive these mechanisms, or are galaxies in groups pre-processed \citep[e.g.][]{Porter2008,Wilman2009,Zabludoff1998a} prior to infall onto the cluster?\\
    
    Studying the neutral gas (\hi) in galaxies can provide an early picture of which environment (group or cluster) plays the strongest role in triggering the morphological evolution of galaxies as they migrate into clusters. The \hi disk extends beyond the stellar disk of the galaxy (particularly in the field), making it the most susceptible to environmental interactions \citep[e.g.][]{Haynes1984a,Serra2013,Jaffe2015}. Strangulation, the process which removes gas from the halo thus halting the fresh supply of \hi to the galaxy, is thought to be one of the main drivers behind the morphology-density relation \citep{Balogh2000,Balogh2000a,VanDenBosch2008}. However, strangulation alone cannot fully explain the transition of galaxies from blue star-forming spiral galaxies to the red, passive lenticular and elliptical galaxies that dominate the galaxy clusters. The effects of galaxy harassment can completely destroy the disks of Sc/Sd galaxies such that the orbits of the constituent stars are affected. It is thought that this could explain the formation of dwarf elliptical (dE) galaxies which are prevalent in the centre of clusters \citep{Mo2010}. Ram-pressure stripping \citep{Gunn1972,Chung2009,Kapferer2009,Abramson2011,Jaffe2015,Yoon2017} is the process by which the gas is stripped from the disk of a galaxy due to its interaction with the intra-cluster or intra-group medium. Ram-pressure stripping is highly effective at rapidly removing the gas from the disks of galaxies, and could be a driving factor in affecting the transition of Sa/Sb to S0s \citep{Mo2010}. Signatures of these different processes affecting the evolution of galaxies are particularly evident in the \hi. Strangulation results in a truncated or anaemic \hi disk, while galaxies that have undergone ram-pressure stripping or harassment tend to have asymmetric \hi disks or tails. The effectiveness of these processes in dense environments means that galaxies in clusters are more \hi deficient than their field counterparts \citep[e.g.][]{Chung2009,Solanes1996}. \\
    
    The Coma cluster (Abell 1656), a nearby rich cluster, is one of the most well-studied galaxy clusters in the Universe \citep[see][for historical review]{Biviano1995}. One of the most comprehensive projects targeting Coma and spanning the entire electromagnetic spectrum, is the Coma Cluster Treasurary Survey\footnote{\url{http://astronomy.swin.edu.au/coma/project-overview.htm}} (CCTS). At the heart of the survey was an observing program using the Advanced Camera for Surveys (ACS) on the Hubble Space Telescope (HST), the goal of the Coma ACS survey was to image the core and the infall region \citep{Carter2008TheDesign}, however the observing program was never completed due to the failure of the ACS. While the HST imaging was never completed, contributions to CCTS from observing programs on other telescopes have provided rich insight into our understanding of the cluster and its population \citep[e.g.][]{Yagi2007TheCluster,Yagi2010,Smith2008}.  \\
   
    At a distance of \numunit{\sim 100}{\text{Mpc}} and velocity dispersion of \numunit{\sim 1000}{\kms}, Coma is the nearest large cluster to us. While Coma is often held up as a large, relaxed cluster, studies of its x-ray emission show a complex dynamical state \citep[see][and references therein]{Neumann2001,Neumann2003}. There are many theories as to the formation of Coma, however the presence and difference in velocities of the two large cD galaxies (NGC 4874 and NGC 4889) at the centre of the cluster suggest that Coma is the result of a merger \citep{Fitchett1987}. Coma has been the target of many redshift surveys in an effort to understand the distribution and kinematics of the galaxies within the cluster and identify any substructure \citep[e.g.][]{Colless1996, Chiboucas2010, SDSSDR13}. The presence of substructure in the cluster is now well established \citep[e.g.][]{Fitchett1987,Mellier1988,Neumann2001,Neumann2003,Adami2005}. Many of the galaxy groups making up the substructure are thought to have been accreted onto the cluster, and in some cases could still be on first infall \citep[e.g.][]{Neumann2001,Neumann2003,Adami2005}. \\
   
    There have also been a number of surveys of the \hi content of galaxies in Coma, both targeted \citep[e.g.][]{Bravo-Alfaro2001, Gavazzi2006} and blind \citep[e.g.][]{Beijersbergen2003, Haynes2018}. These works have provided insight into the ongoing gas removal processes affecting the galaxies in the cluster, as well as an understanding of the \hi deficiency of the cluster. The previous Coma \hi surveys have lacked coverage, sensitivity, resolution, or some combination thereof. In this work, we use \hi observations from the blind, high sensitivity and high resolution survey with the Westerbork Synthesis Radio Telescope (WSRT): the Westerbork Coma Survey (WCS, Molnar et al. in prep). With the wealth of information already available, in combination with new \hi observations of the cluster, we can explore the morphologies of the Coma galaxies in conjunction with their gas properties and local environment. Coma is known to be home to a large number of S0 galaxies: the relative fraction of S0s, ellipticals, spirals and mergers is \numunit{42}{\text{per cent}}, \numunit{22}{\text{per cent}}, \numunit{32}{\text{per cent}}, and \numunit{4}{\text{per cent}} respectively \citep{Beijersbergen2003}. This makes Coma an ideal laboratory to study \hi properties of galaxies of different morphologies in the different local environments. \\
   
    In this paper, we make use of data from a blind \hi survey of the Coma cluster (\secref{sec:coma_wcs}) from \cittmolnar to study the \hi gas properties of the different galaxy types in different environments within the cluster. Most of the galaxies in Coma are not individually detected in \hi, thus we make use of the \hi stacking technique (\secref{sec:coma_histackingmethod}) to improve the signal-to-noise ratios of groups of \hi spectra. The optical spectroscopy and photometry used in this work are presented in \secref{sec:coma_ancdata}. We explore the global \hi properties of the galaxies in Coma by deriving optical-\hi scaling relations and by determining the \hi deficiency at different locations within the cluster (\secref{sec:coma_globalhi}). We use the substructure within the cluster as a measure of the local environment (\secref{sec:coma_substructure}), and use \hi stacking to probe the \hi content of samples of galaxies within local environments relative to the global \hi properties of the cluster (\secref{sec:coma_locaglobal}). {Throughout this paper we use \numunit{\mathrm{H}_0 = 70}{\text{km}\,\text{s}^{-1}\,\text{Mpc}^{-1}} ($h = 0.7$), $\Omega_m = 0.7$, and $\Omega_\Lambda = 0.3$.}

\section{Multi-wavelength data}
    \label{sec:coma_ancdata}
    \subsection{Optical Spectroscopy}
    \label{sec:coma_spectroscopy}

    A full, wide-area redshift census of Coma is essential to this work for two reasons: 1 -- \hi stacking requires spectroscopic redshifts in order to align the \hi non-detected spectra; 2 -- we require redshifts in order to more accurately characterise the environment (sub-structure) within the cluster, as well any infalling groups. This work makes use of redshifts from two different sources: literature (\secref{sec:coma_litredshifts}) and spectroscopy obtained with Hydra at the WIYN Telescope\footnote{The WIYN Observatory is a joint facility of the University of Wisconsin-Madison, Indiana University, the National Optical Astronomy Observatory and the University of Missouri.} (\secref{sec:coma_wiynredshifts}). The final catalogue contains 1095 spectroscopically confirmed members of the Coma cluster within a \numunit{2}{\text{deg}} radius of the cluster centre (58 are new redshifts -- see \secref{sec:coma_wiynredshifts}); 850 of these fall within the Westerbork Coma Survey footprint (the orange outline in \figref{fig:comatargets}). The larger catalogue is necessary when looking for sub-structure within the cluster.

    \subsubsection{Redshifts from the literature}
        \label{sec:coma_litredshifts}
          
          The literature catalogue was collated from a number of published and unpublished catalogues. More than two thirds of the redshifts were obtained from the SDSS Data Release 13 \citep[DR13;][]{SDSSDR13}. The remainder come from a number of different sources (see \tabref{tab:litredshiftlist} for the number of redshifts from the different sources) -- the reference for each redshift is given in the Coma galaxy catalogue in Appendix C. \\
          
          Using the VizieR astronomical catalogue database \citep{Ochsenbein2000}, we queried data in published catalogues within a \numunit{2}{\text{deg}} radius centred on the Coma cluster. Merging the SDSS redshifts with those obtained from VizierR created a large master catalogue with overlapping sources, some of which had redshifts outside of the Coma volume. To construct our catalogue of unique sources associated with the Coma cluster, we only consider sources with \numunit{3000 < cz < 10500}{\kms} as this is the spectral range spanned by the \hi data set. Sources within $5''$ radius of each other are considered to be the same source; when multiple sources within the search radius were found, only the redshift with the smallest uncertainty was kept in the catalogue. Every galaxy with multiple redshift measurements was checked by eye so as not to remove the redshifts of real spatial overlapping galaxies.

    \subsubsection{Spectroscopy with Hydra}
      \label{sec:coma_wiynredshifts}
    
        \begin{figure}[h]
            \centering
            \includegraphics[width=\linewidth]{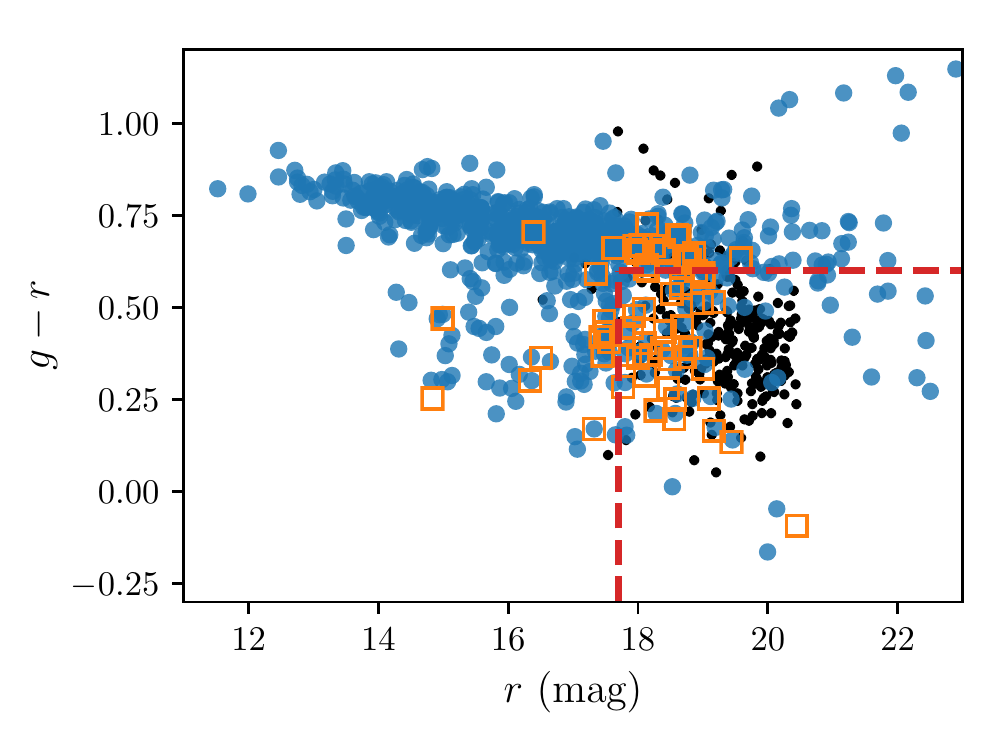}
            \caption{Colour magnitude diagram for the Coma cluster. The blue points represent the galaxies in the catalogue of literature redshifts. {The black dots represent all galaxies targeted with Hydra, but whose redshifts place them in the foreground or background. The orange open squares represent the galaxies targeted with Hydra with redshifts that place them inside the cluster.} Galaxies with \numunit{r > 17.7}{\text{mag}} and $\gmr < 0.6$ were prioritised {as indicated by the dashed red lines.}}
            \label{fig:coma_targetcmd}
        \end{figure}

        \begin{figure*}[h]
          \centering
          \includegraphics[width=\linewidth]{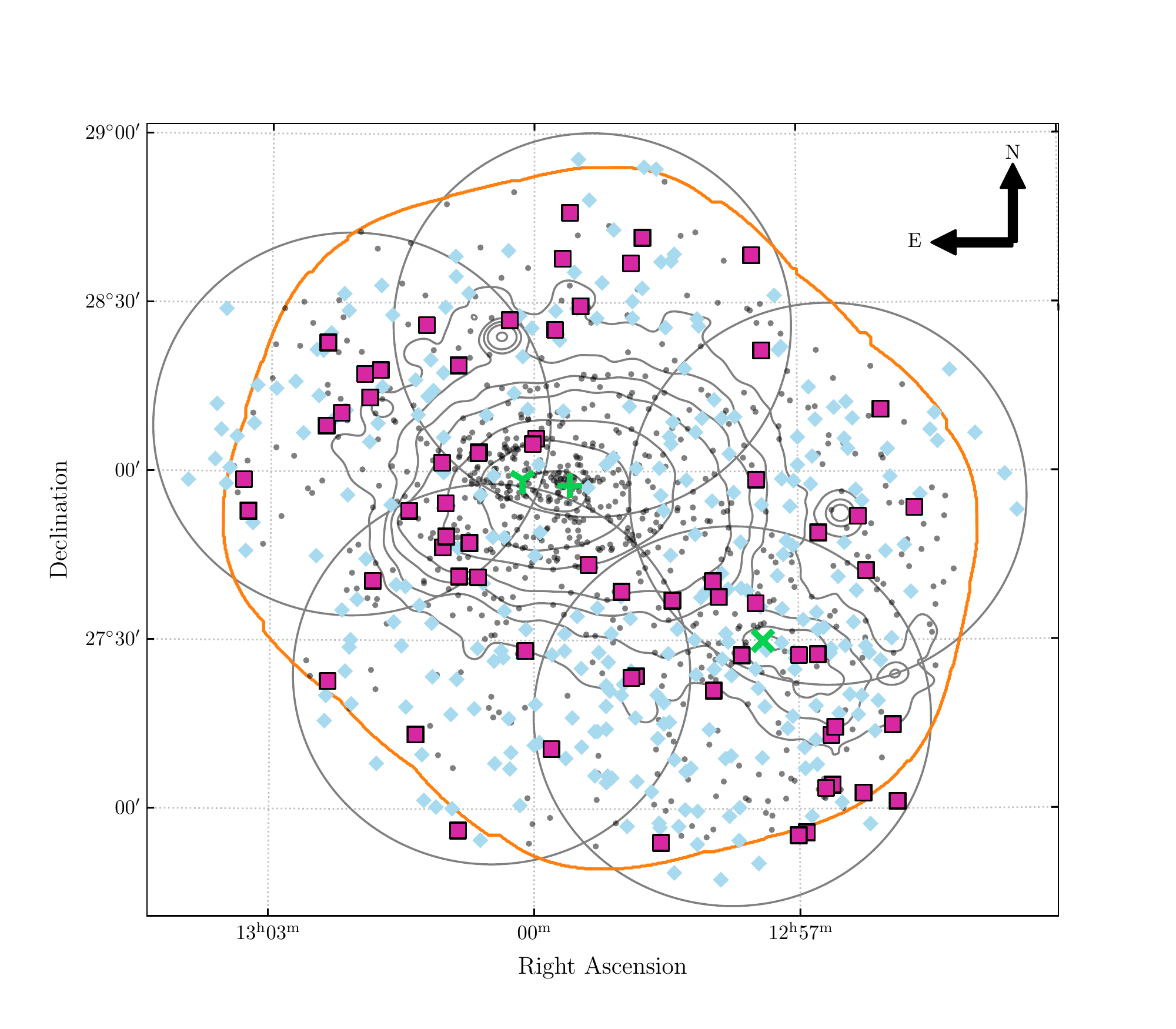}
          \caption{WIYN/Hydra targets, and observed sources. The five fields (3 of the fields had a second configuration) targeted with WIYN are indicated by the large circles. The square and diamond data points represent the galaxies which we targeted with Hydra. The galaxies with redshifts in the Coma volume are represented by the magenta square data points and foreground/background galaxies are represented by the blue diamond markers. The green markers represent the three cD galaxies in Coma ($+$--NGC 4874, $\curlyvee$--NGC 4889 and $\times$--NGC 4839). The \numunit{0.4-2.4}{\text{keV}} ROSAT x-ray contours are given by the grey contours (outer contours correspond to \numunit{3\times 10^{-5}}{\text{cts/s}}, and inner-most contours to \numunit{1\times 10^{-3}}{\text{cts/s}}). Other confirmed members of the Coma cluster are represented by the small grey circles. The Westerbork Coma Survey footprint is indicated by the orange outline. }
          \label{fig:comatargets}
        \end{figure*}

        \begin{figure}[h]
          \centering
          \begin{subfigure}{\linewidth}
            \includegraphics[width=\textwidth]{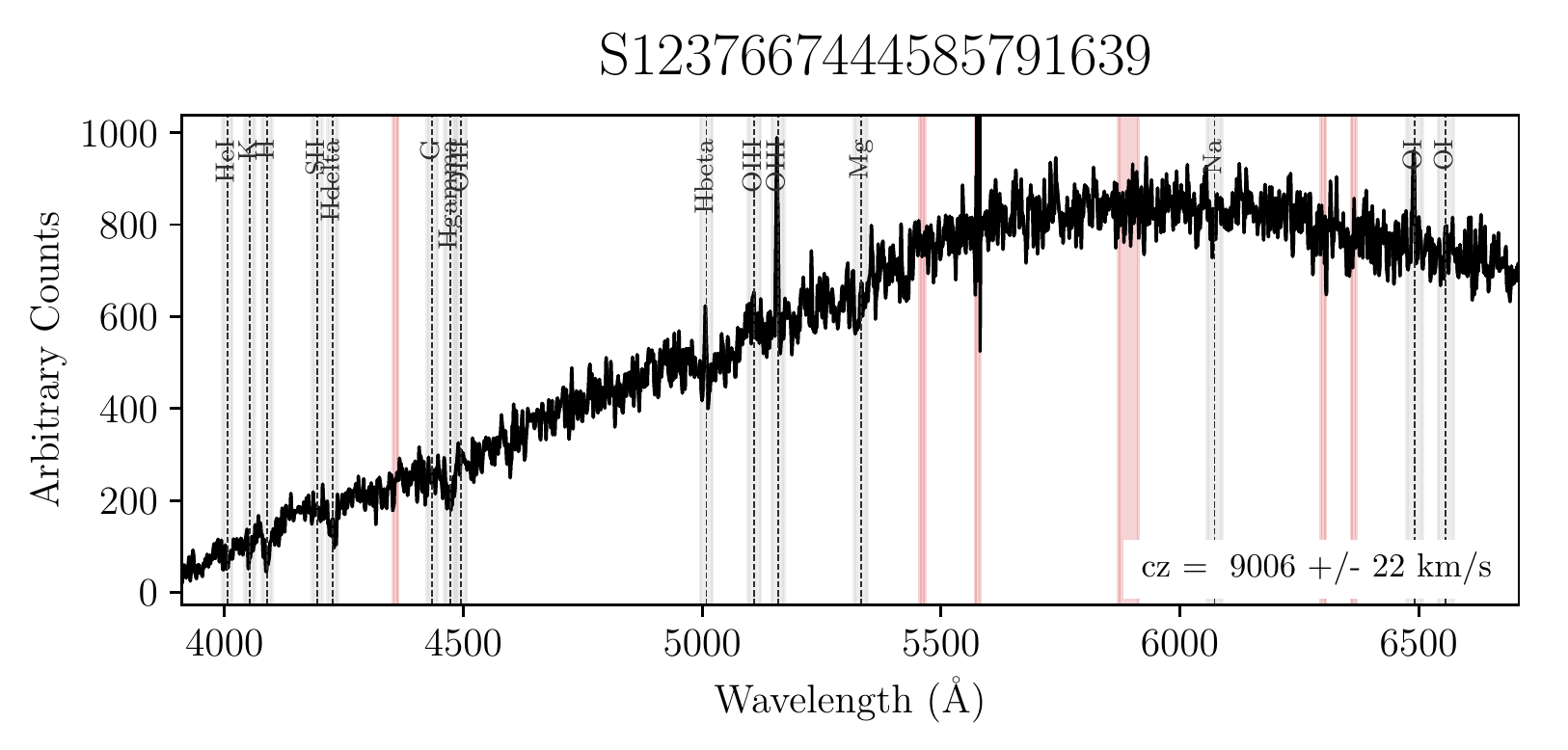}
            \caption{Good spectrum}
            \label{fig:specexample1}
          \end{subfigure}
          \begin{subfigure}{\linewidth}
            \includegraphics[width=\textwidth]{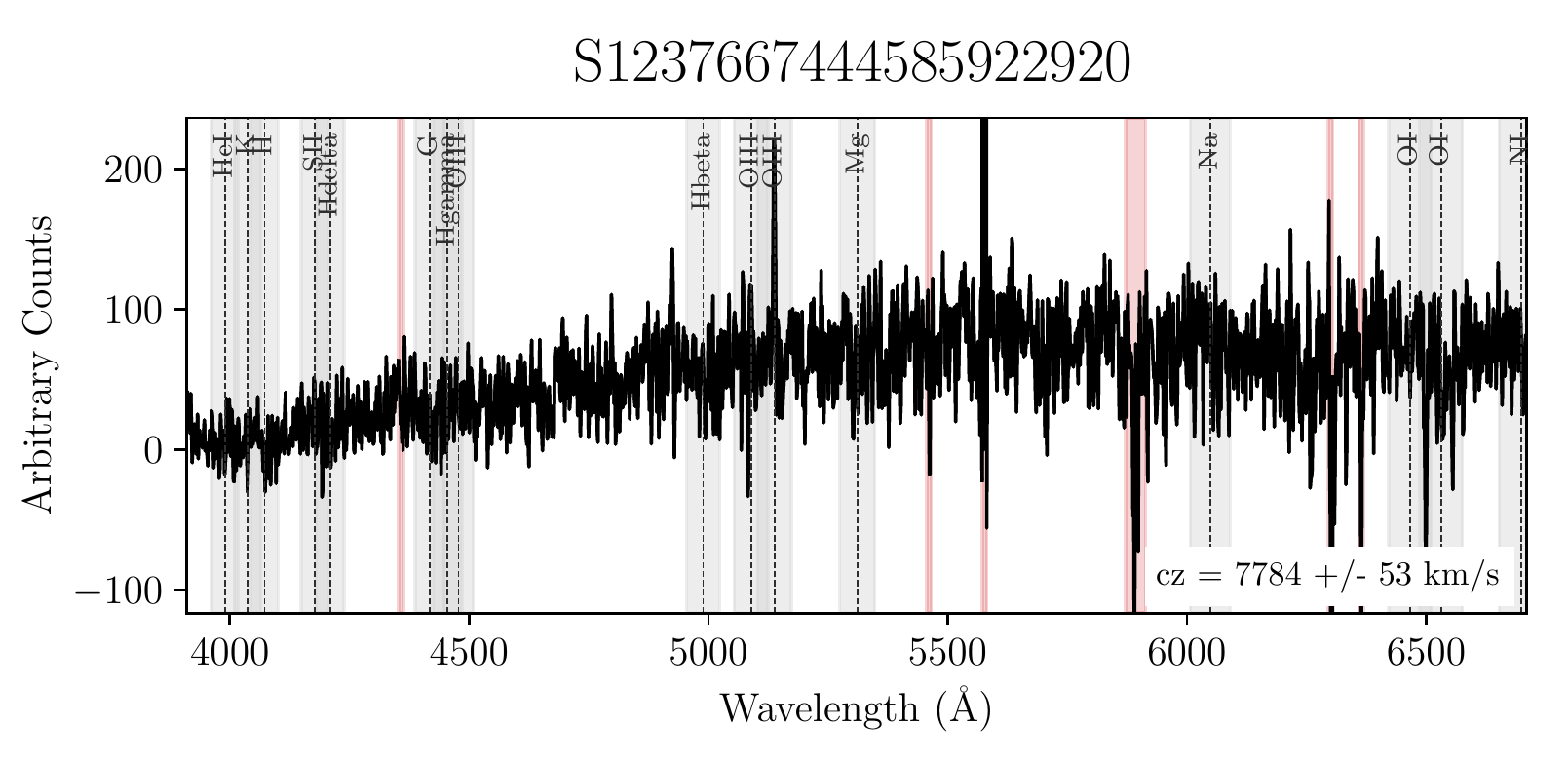}
            \caption{Low signal-to-noise}
            \label{fig:specexample2}
          \end{subfigure}
          \begin{subfigure}{\linewidth}
            \includegraphics[width=\textwidth]{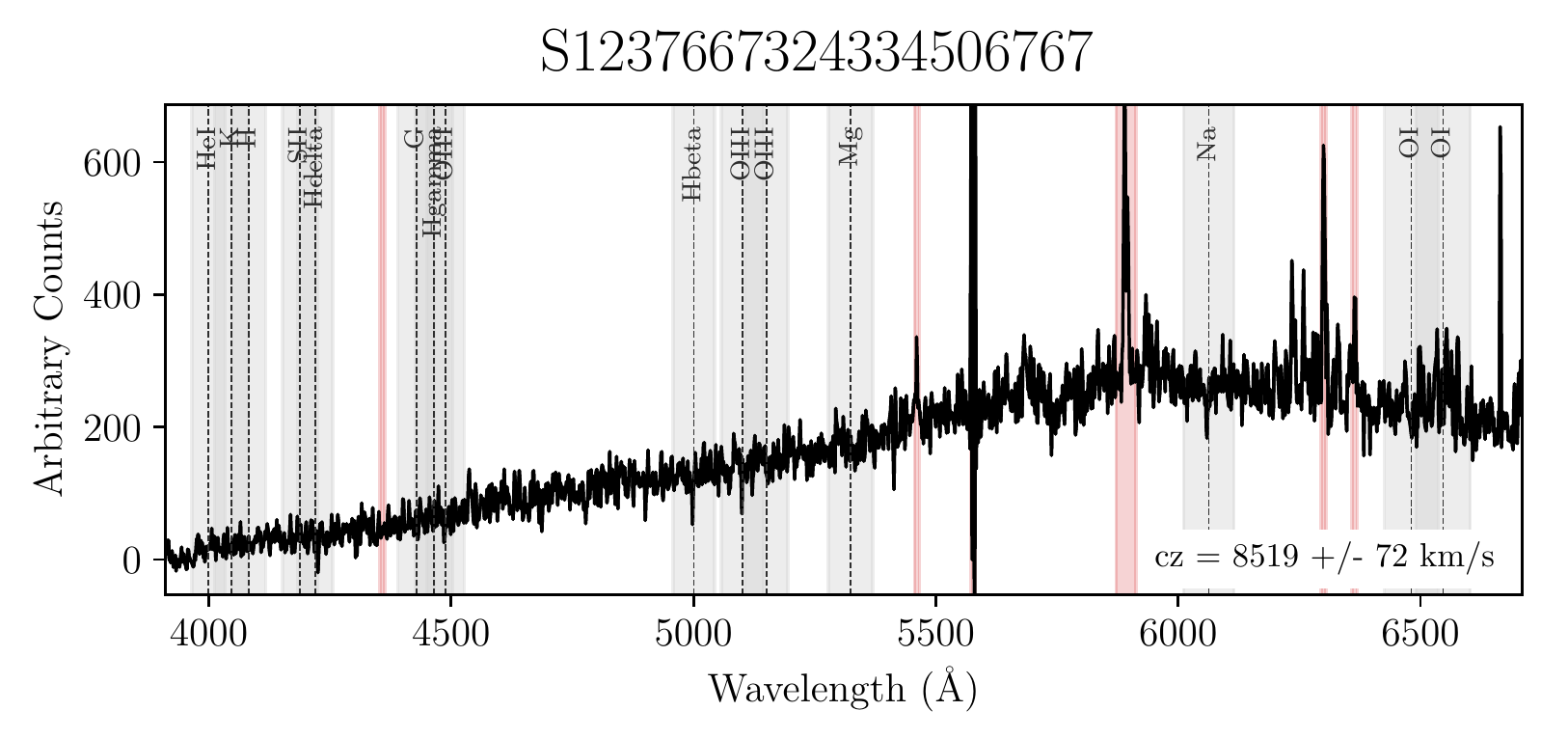}
            \caption{Poor sky subtraction}
            \label{fig:specexample3}
          \end{subfigure}
          \caption[Three examples of the cross-correlation results of Hydra spectra]{Three examples of the cross-correlation results of Hydra spectra. The dashed lines and grey regions indicate the central wavelength and associated error of the labelled line. Red regions indicate those masked out during the cross-correlation process because of strong sky lines. }
          \label{fig:xcexample}
        \end{figure}
        
        New spectroscopic data for galaxies in the Coma cluster was obtained using Hydra, the multi-object spectrograph, on the 3.5m WIYN Telescope located at the Kitt Peak National Observatory. Despite being an extremely well-observed cluster, the spectroscopic coverage of Coma drops off for galaxies fainter than \numunit{r \sim 17.7}{\text{mag}}, thus we targeted galaxies fainter than this limit. Since previous targeted spectroscopic surveys have focused on the core of the cluster, we prioritised galaxies with \numunit{r > 17.7}{\text{mag}} on the outskirts of the cluster and with bluer colours ($g - r < 0.6$) as these are the galaxies which we expect to contain \hi. \figref{fig:coma_targetcmd} shows the colour magnitude diagram (CMD) for the cluster. The lower right quadrant of the CMD shows very few redshifts from literature (indicated by the blue points), another motivation for targeting galaxies in that quadrant of the CMD.  Initially 8 configurations of the 5 fields were targeted (3 of the fields had a second configuration, which was necessary due to the density of targets); however due to poor weather conditions on the first night, only 6 configurations were actually observed. The data were collected on the nights of 19 -- 20 April 2018.  \\ 

        All the data reduction was done using \iraf\footnote{IRAF is distributed by the National Optical Astronomy Observatory, which is operated by the Association of Universities for Research in Astronomy (AURA) under a cooperative agreement with the National Science Foundation.}. The first phase of the data reduction after cosmic ray removal using \texttt{lacosmic} \citep{VanDokkum2001} was done using the \texttt{ccdred} package; this took care of the overscan, bias, and dark subtraction. The second phase was guided by the \texttt{dohydra} task which takes care of the flat-fielding, aperture extraction, wavelength calibration, dispersion correction, and sky subtraction. The wavelength-calibrated and sky-subtracted spectra were then matched to template spectra in order to determine the redshift of the galaxies. \\

\begin{figure}[t]
          \centering
          \includegraphics[width=\linewidth]{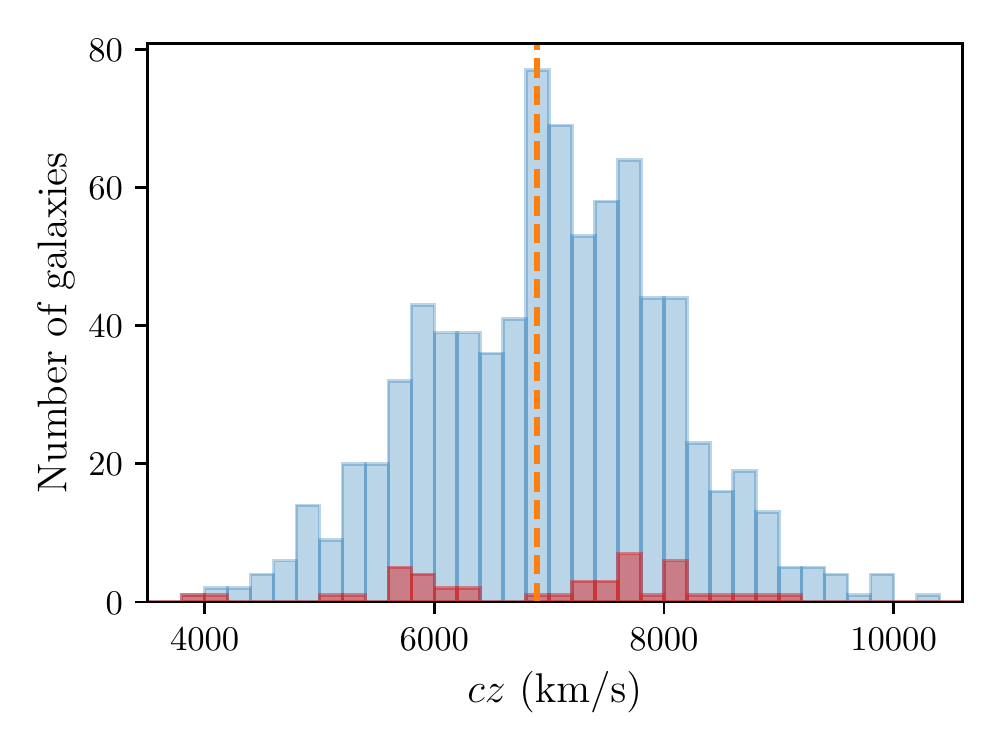}
          \caption[Redshift Histogram]{Distribution of all redshifts (literature and new) within Westerbork Coma Survey footprint. The red represents the galaxies whose redshifts were obtained using Hydra/WIYN.}
          \label{fig:comaredshifthists}
        \end{figure}

        The redshift calculations were done using the cross-correlation technique which is implemented by the \iraf task \texttt{xcsao} \citep{Kurtz1992} contained within the \texttt{rvsao} package. We searched for velocities between \numunit{0}{\kms} and \numunit{100\,000}{\kms} using seven different templates. {We used three different types of emission line templates, three different types of absorption line templates and one composite template so as to make sure we had covered the range in different types of stellar populations.} Velocities were only accepted as robust if at least 3 templates produced results within \numunit{100}{\kms}, a common uncertainty for a measurement based on absorption lines. The uncertainty on each redshift measurement was determined from the peak of the correlation between the template and observed spectrum. A detailed explanation can be found in \citet{Kurtz1992}. All results were also inspected by eye.
        Three examples of Hydra spectra are shown in \figref{fig:xcexample}. The spectra have been sky-subtracted, but have not been flux calibrated as this is not necessary to obtain a redshift measurement; the vertical red bands indicated the regions of the spectra where there are strong sky lines, the dashed lines and grey regions indicate the wavelength and associated error of the labelled lines. Of the 363 observed galaxies, the spectra of 278 galaxies had high enough signal-to-noise to obtain a redshift measurement (see \figref{fig:specexample1}); 58 of these redshifts are new measurements of galaxies in the Coma cluster (see the red histogram in \figref{fig:comaredshifthists} for redshift distribution of these galaxies). The 95 galaxy spectra from which no redshift measurement could be made, were discarded from the sample due to low signal-to-noise (see \figref{fig:specexample2}) or poor sky subtraction (there were 29 such galaxy spectra -- see \figref{fig:specexample3}). The poor sky subtraction is evident in the spectrum shown in \figref{fig:specexample3} by the ``emission line'' features in the red vertical bands, these bands indicate the regions where there are known strong sky lines. 

  \subsection{Photometric Data}
    \label{sec:coma_photometry}
       
      Despite being an extremely well-studied cluster, there are no publicly available matched-aperture or total magnitude photometric catalogues that cover the entire region of Coma that we are interested in. It is important when comparing flux (or magnitude) measurements in different photometric bands that the measurements are determined in the same manner (i.e. matched apertures or model fits for total magnitudes). The overlapping sources at the centre of the cluster also provide a challenge to many aperture-based photometric pipelines. Our imaging data includes near- and far-UV data from GALEX \citep{Bianchi2000}, five optical bands (\textit{ugriz}) from SDSS \citep[DR12][]{Alam2015}, as well as the four mid-infrared bands (W1 - W4) from WISE \citep{Wright2010}. Frames from the 11 bands were mosaicked using \textsc{Swarp} \citep{Bertin2010} to create a single \numunit{2 \times 2}{\text{deg}^2} image per band centred on the cluster, all with the same pixel scale (\numunit{0.396}{''/\text{px}}). We also created a pseudo B-band image from the SDSS \textit{g} and \textit{i} bands using the following relation from \citet{Cook2014} for conversion from SDSS to Johnsons magnitude:
      \begin{equation}\label{eqn:bband}
        B - i = (1.27 \pm 0.03) ( g - i ) + (0.16 \pm 0.01).
      \end{equation}
      This relation was calibrated using galaxies in the Local Volume Legacy Survey \citep{Dale2009}. It should be noted that this pseudo B-band magnitude is only used to calculate the diameter ($D_{25}$) of each galaxy at the \numunit{25}{\text{mag/arcsec}^2} isophote which is required for estimating expected \hi mass (see \secref{sec:coma_globalhi}). \\

      In order to calculate magnitudes and other morphological properties for each galaxy for which we have a redshift measurement within the WCS footprint, we follow the strategy used by \citet{Barden2012} in the \textsc{Galapagos} pipeline. For each galaxy we create a set of postage stamps that are then used to calculate properties such as magnitude and size. The modelling is done using \textsc{GalfitM} \citep{Bamford2011} which is a modified version of \textsc{Galfit} created by \citet{Peng2002,Peng2010}. \textsc{GalfitM} is capable of modelling a source across multiple bands simultaneously. It is very sensitive to the initial parameters, and so we therefore calculate the initial values using source characterisation tools in \texttt{photutils} (an \texttt{AstroPy} package for photometry). We also use the \texttt{photutils} image segmentation feature to create masks for each band: any source not associated with our target is masked, with the exception of overlapping sources which are modelled simultaneously using \textsc{GalfitM}. There are 23 galaxies where the fit failed or was not possible due to the galaxy being too faint (four of these are known ultra-diffuse galaxies). These galaxies are still included in our spectroscopic sample as they come from spectroscopic surveys that targeted extremely faint dwarf galaxies \citep[e.g.][]{Chiboucas2010} or ultra-diffuse galaxies \citep[e.g.][]{Alabi2018}.\\

      \begin{figure*}[h]
        \centering
        \includegraphics[width=\linewidth]{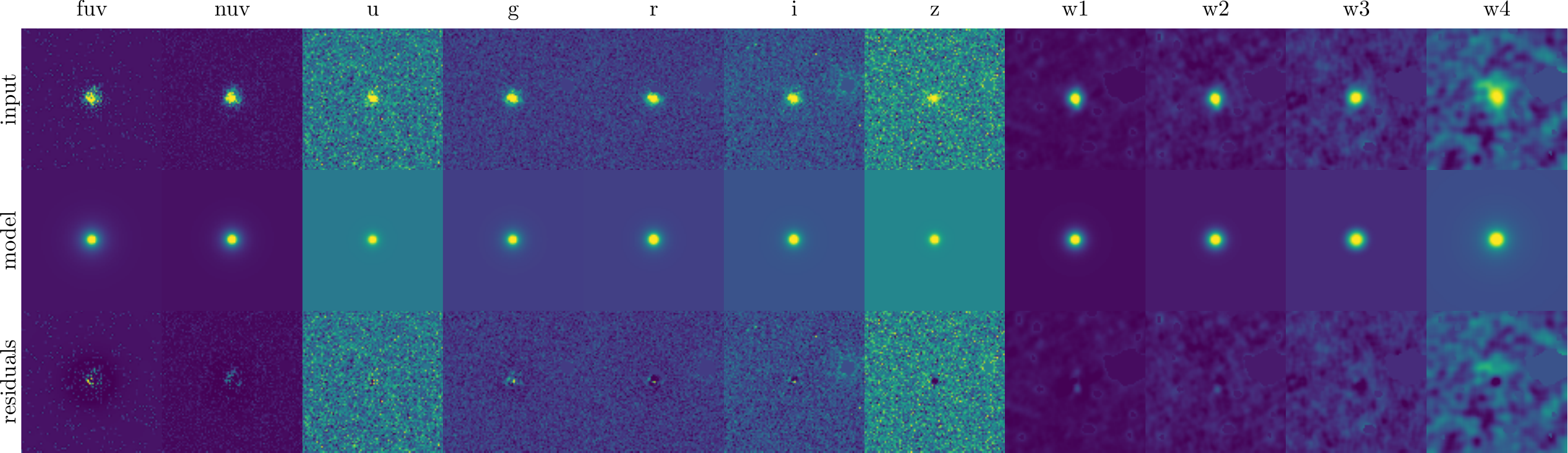}
        \caption[Example of galaxy modelling with \textsc{GalfitM}]{A galaxy modelled with a single Sersic profile by \textsc{GalfitM}. The input data are shown in the top row, the model in the second row and the residuals in the bottom row. The colour scale is set in every column such that each row has the same intensity.}
        \label{fig:galfitm_example}
      \end{figure*}

      An example of a single Coma galaxy modelled using a single Sersic profile by \textsc{GalfitM} is presented in \figref{fig:galfitm_example}. The Sersic profile, one of the most commonly used profiles for modelling galaxies of differing morphologies \citep{Peng2010}, is given by
      \begin{equation} \label{eqn:sersic}
          I(r) = I_e \exp\left[  -\kappa \left( \left(\frac{r}{r_e}\right)^{1/n} - 1 \right) \right]
      \end{equation}
      where $I_e$ is the pixel intensity at the effective radius radius ($r_e$). The $\kappa$ is tied to the value of $n$ such that when $n = 4$ (making \autoref{eqn:sersic} consistent with a de Vaucouleurs profile), $\kappa = 7.67$ \citep{Peng2010}. Galaxies better described by an exponential profile, will have $n = 1$. During the model fitting process, we allow the model parameters to vary across the bands. The full catalogue of properties for Coma galaxies withing the WCS footprint is given in Appendix C.  \\

      \begin{figure*}[h]
        \centering
        \includegraphics[width=\linewidth]{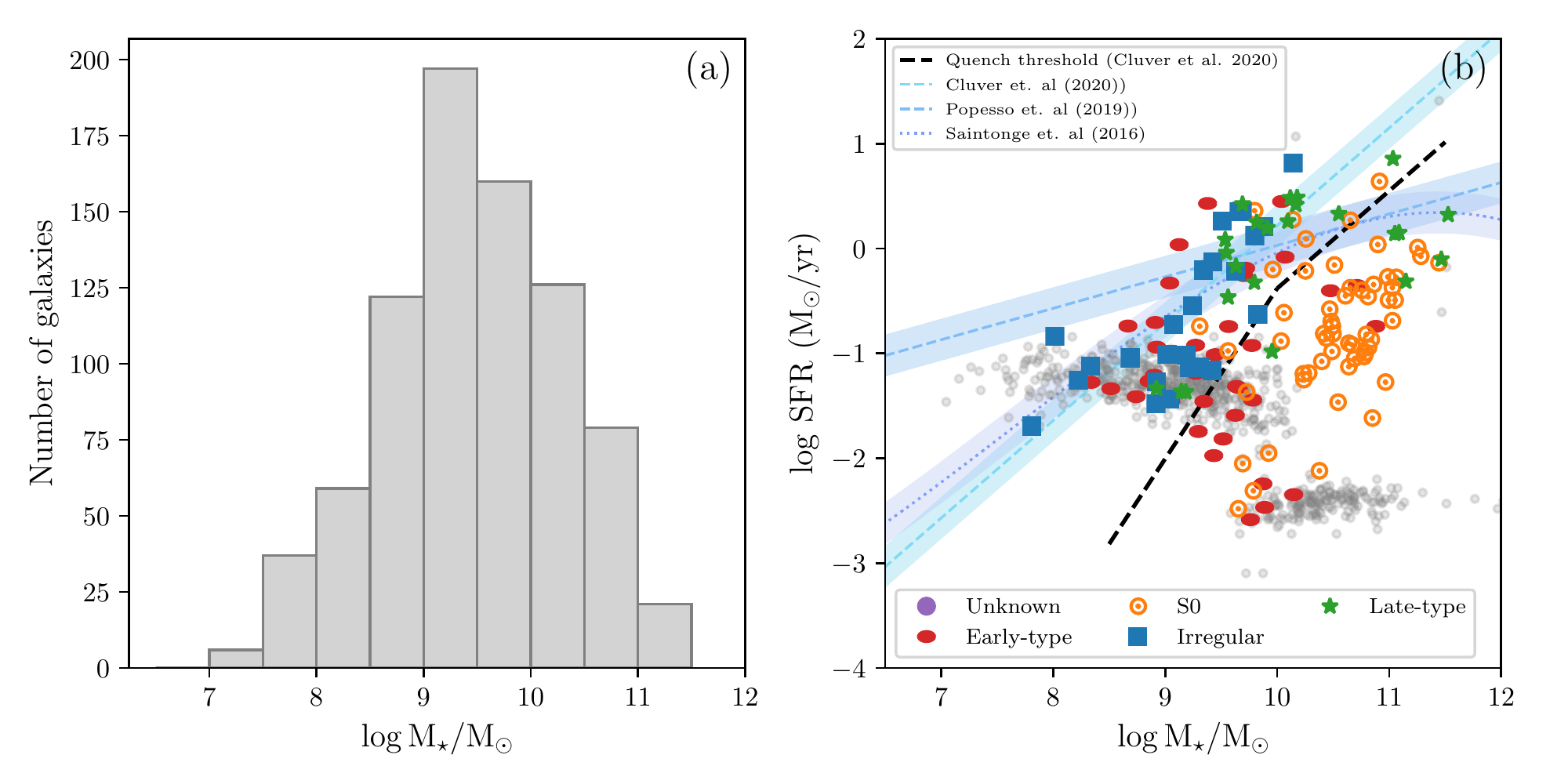}
        \caption[Stellar mass and star formation rate distribution]{\textit{Left:} stellar mass distribution of Coma galaxies within the WCS footprint. \textit{Right:} star formation rate as a function of stellar mass for the Coma galaxies. The $2\sigma$ upper limit on the star formation is shown for galaxies where there is no detection in the W3 band, these are represented by the light grey points. The step function in the upper limits arises due to a non-detection in W3 for low mass galaxies and a non-detection after continuum subtraction in W3 for high mass galaxies. Also shown for reference are fits to the star formation main sequence by \citet{Cluver2020,Popesso2019,Saintonge2016}. The black dashed line indicates the {quench threshold \citep{Cluver2020}} below which galaxies are not considered to be actively star-forming.}
        \label{fig:coma_smass_sfr}
      \end{figure*}

      Stellar masses and star formation rates for the Coma galaxies are calculated following the method outlined in \citet{Cluver2014} and using a custom pipeline \citep{Jarrett2013,Jarrett2019}. The mass-to-light ratios for each galaxy are calculated from the WISE W1-W2 colour using
      \begin{equation}\label{eqn:wisemstar}
        \log \mathrm{M}/L_\mathrm{W1} = -1.96(\mathrm{W1 - W2}) - 0.03,
      \end{equation}
      where $L_\mathrm{W1}$ is the W1 (\numunit{3.4}{\mu\text{m}}) inband luminosity of the galaxy relative to the sun, and ($\mathrm{W1 - W2}$) is the rest-frame colour \citep{Cluver2014}. The distribution of stellar masses for Coma is shown in panel (a) of \figref{fig:coma_smass_sfr}. \\

      As has been shown by \citet{Cluver2014,Cluver2017}, the W3 (\numunit{12}{\mu\text{m}}) and W4 (\numunit{23}{\mu\text{m}}) bands trace the interstellar medium emission (ISM) in galaxies and can thus be used to determine star formation rates (SFR). The SFRs used in this work are calculated from the \numunit{12}{\mu\text{m}} luminosity (W3) relation determined by \citet{Cluver2017}. The W3 flux measurements are separated into the contributions from polycyclic aromatic hydrocarbon (PAH) emission (which is related to the star formation activity) and a contribution from the stellar continuum \citep{Cluver2017}. Panel (b) of \figref{fig:coma_smass_sfr} shows the star formation as a function of stellar mass for the sample. Many of the Coma galaxies do not have a measured SFR, and so are presented as $2\sigma$ upper limits in \figref{fig:coma_smass_sfr}. There is a step-distribution in the SFR upper limits which arises due to the low mass galaxies (\numunit{\mstar \lesssim 10^{9.5}}{\msol}) not detected in the W3 band. The higher mass galaxies (\numunit{\mstar \gtrsim 10^{9.5}}{\msol}) are detected, but the W3 flux is largely attributed to the stellar continuum which is removed prior to determining the SFR.

  \subsection{H {\sc i} data}
    \label{sec:coma_wcs}

    The blind, high sensitivity and high resolution WSRT \hi observations from \cittmolnar provide an opportunity to study the gas content of Coma galaxies located well within the cluster environs, and those on infall. Like most dense galaxy environments, Coma is well-known for being \hi deficient, and so we use the \hi stacking method to probe the average \hi content of groups of galaxies. \\
 
    The WCS observed Coma in two overlapping \numunit{20}{\text{MHz}} bands (\numunit{1369-1389}{\text{MHz}} and \numunit{1386-1406}{\text{MHz}}) and is comprised of 24 overlapping WSRT pointings of \numunit{1 \times 12}{\text{h}}, making the effective integration, due to the overlap, \numunit{3 \times 12}{\text{h}} per pointing. The final data cubes were created using the MIRIAD software package \citep{Sault2011}. For details about the data reduction see \citet[in prep.]{Serra2019}. The two data cubes, together, cover approximately \numunit{\sim 5}{\unit{deg}{2}} and a velocity range of \numunit{\sim 3000 - 10500}{\kms}, however, there is a break in velocity range of \numunit{\sim 24}{\kms} at the central cluster velocity of \numunit{6925}{\kms}. The two \hi cubes have a spatial resolution of \numunit{26.5 \times 40.3}{\unit{arcsec}{2}} and a velocity resolution of \numunit{16.5}{\kms}. \\
    
    The WCS contains 39 galaxies for which there are direct \hi detections, with \hi masses greater than \numunit{4.4 \times 10^7}{\msol}. The detections were identified using SoFiA \citep{Serra2015} and confirmed by eye. For a details on the source finding and discussion about the properties of the \hi detections, see \cittmolnar.

\section{H {\sc i} Stacking Method}
    \label{sec:coma_histackingmethod}
    
  The technique utilized in this study to probe the \hi content of the galaxies in the Coma cluster is called \hi stacking. This is a statistical technique that has become popular in order to push the \hi mass sensitivity below the observation limit. The technique is particularly useful for studying the average \hi properties of samples of galaxies which are predominantly not directly detected in \hi. \hi stacking has been successfully used to study the neutral gas content of galaxies in clusters and dense environments \citep[e.g.][]{Brown2016c, Chengalur2001, Fabello2012, Jaffe2016, Lah2009, Verheijen2007}. \\

  \hi stacking uses the optical redshift information for each galaxy to align the \hi spectra. The aligned spectra are co-added to create an average spectrum with lower noise statistics than the individual spectra. In this study we make use of the publicly available \hi Stacking Software \citep[HISS,][]{Healy2019}\footnote{\url{github.com/healytwin1/HISS}} tool to perform all our stacking experiments. \\

  For every galaxy with a redshift in our catalogue, we extract a spectrum from the \hi data cubes. For galaxies where the B-band $D_{25}$ is resolved by the beam, we extract the spectrum using an aperture with the semi-major axis determined from the galaxy diameter at the B-band \numunit{25^{th}}{\text{mag}} isophote ($D_{25}$) (see \figref{fig:spectrumextract_res}). For galaxies unresolved by the beam, the corresponding \hi spectra are extracted using an aperture the size of the beam ($25'' \times 40''$) (see \figref{fig:spectrumextract_unres}). Along with the target spectrum, for every galaxy, we also extract 25 ``reference'' spectra using the same aperture as the target spectrum. The bottom left panel of \figref{fig:spectrumextract_unres} and \figref{fig:spectrumextract_res} show the locations around the target from which the 25 reference spectra are extracted. These reference spectra are stacked in all our analyses in the same manner as the target galaxy spectra. For each target galaxy we extract background spectra from 25 offset positions around it. Taking as input a background spectrum from the same relative offset position around each target galaxy, we produce a background reference spectrum. Using all the offset positions results in 25 separate reference spectra per stacked target spectrum, with each reference spectrum containing the same number of stacked spectra as the target stack. Stacked reference spectra are used to aid in quantifying the signal in the main stacked spectrum.The average of the 25 stacked reference spectra for a well calibrated dataset should be around zero (e.g. see the grey line in the panels of \figref{fig:coma_phase_stacks}). From the spread of the 25 stacked reference spectra, we estimate the uncertainty on the quantity measured from the stacked target spectrum.
    
 \section{H {\sc i} in Coma}

        \label{sec:coma_globalhi} 
            
   \subsection{H {\sc i} -- \mstar scaling relation}

\begin{figure*}[h]
      \centering
      \includegraphics[width=\linewidth]{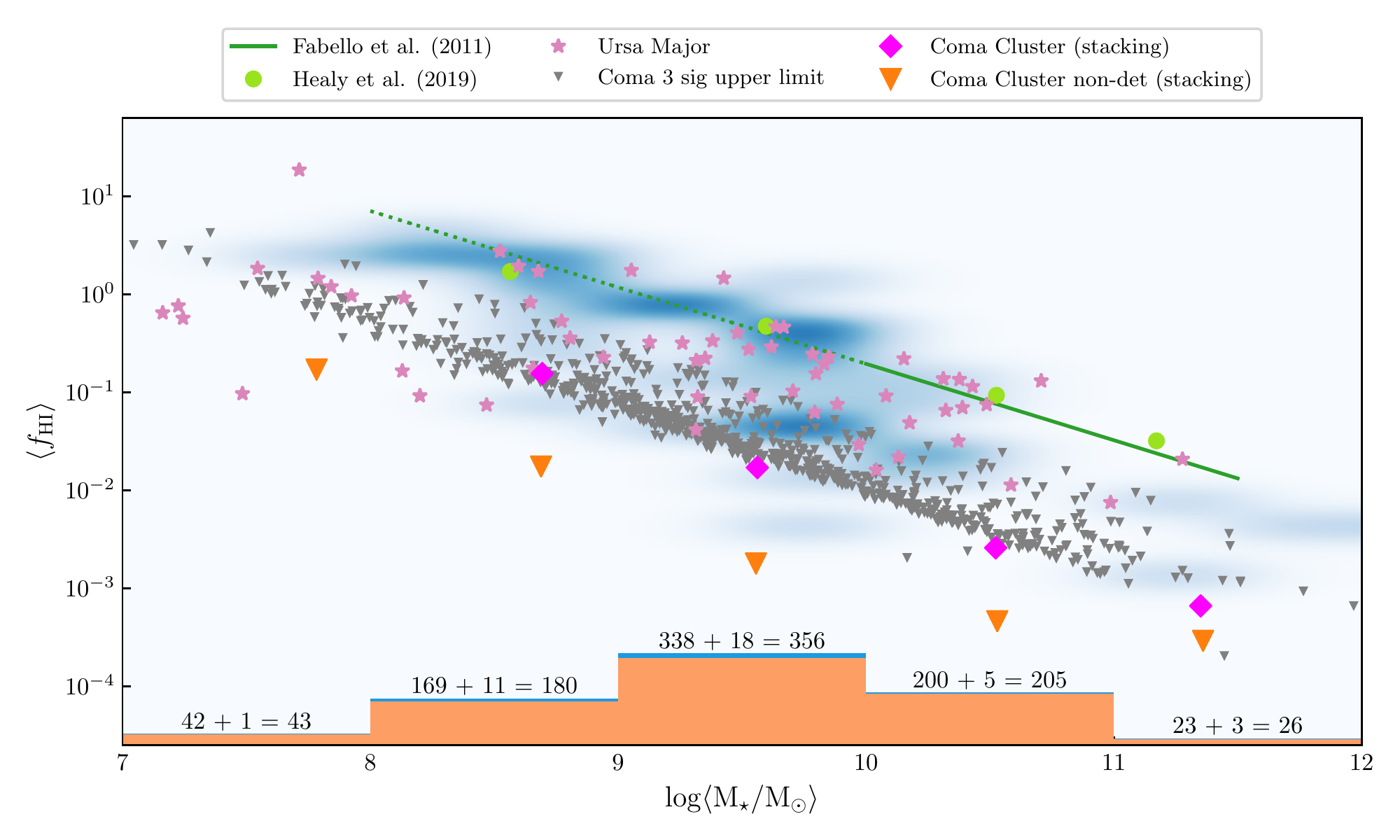}
      \caption[Coma \mhi to \mstar scaling relation]{Stellar mass (\mstar) to \hi gas fraction (\fhi) scaling relation for the Coma cluster. The individual $3\sigma$ upper limits for the Coma \hi non-detections are represented by the grey lower triangles while the distribution of the individual \hi detections is shown by the blue density plot. The pink diamond and orange lower triangle symbols represent the stacking of all Coma galaxies and only non-detections respectively. {The pink diamonds show measurements from a detected stacked spectrum, and the orange triangles indicate the $3\sigma$ upper limit in the case of a non-detection in the stacked spectrum.} The round light green symbols and green line represent the stacking of field galaxies by \citet{Healy2019} and \citet{Fabello2011} respectively{, the dashed green line is an extrapolation of the \citet{Fabello2011} results}. For reference, we also show the individual \fhi measurements (pink stars) for galaxies in the Ursa Major cluster \citep{Verheijen2001a}. }
      \label{fig:coma_gasfractions}
    \end{figure*}

    The stellar mass (\mstar) to \hi gas fraction (\fhi, defined as the \hi mass to stellar mass fraction) scaling relations (\mstar-\fhi) are well characterised for galaxies in the field \citep[e.g.][]{Fabello2011,Brown2015,Healy2019}, and have also been used to study the deficiency of \hi in galaxies in different density environments including the Virgo cluster \citep[e.g.][]{Cortese2011,Fabello2012}. However, given the small number of direct \hi detections in Coma, there has been no study of the \mstar-\fhi scaling relation as a method to probe the \hi content of the cluster as a whole. In \figref{fig:coma_gasfractions}, we present the \mstar-\fhi scaling relation for Coma. \\
    
    We separate the galaxies into samples by stellar mass (indicated by the histogram in \figref{fig:coma_gasfractions}) to probe the \fhi-\mstar scaling relation. \figref{fig:coma_gasfractions} shows the Coma direct \hi detections in the blue distribution and the \hi mass upper limits for the \hi non-detections in grey lower triangles. The \hi mass upper limits are calculated using the noise from the individual spectra and an upper estimate on the $W_{50}$ line width calculated using the $r$-band Tully Fisher relation from \citet{Ponomareva2018}. The scatter in the distribution of upper limits can also be attributed to the spread in the noise distribution of the WCS mosaic. Given the number of galaxies that are not detected in \hi, we use the \hi stacking technique to probe the \hi content below the detection threshold. \\
    
    Utilizing the \hi stacking technique, we were able to improve the sensitivity in the stacked spectrum by more than order of magnitude relative to the individual spectra. We stack both detections and non-detections, as well as just non-detections. Despite the improved sensitivity of the stacked spectra, there are no detections in the stacked spectra corresponding to the sample with only non-detections (the stacked spectra for each stellar mass bin can be found in Figs.~\ref{fig:coma_stacks_mstarfc}~\&~\ref{fig:coma_stacks_mstarnd}). The $3\sigma$ upper limits for the stacking results (indicated by the orange triangles) suggest that Coma galaxies on average have at least three orders of magnitude less gas than galaxies at the same stellar mass in the field (indicated by the round light green symbols \citep{Healy2019} and the dark green line \citep{Fabello2011}). In comparison, galaxies in Ursa Major cluster (indicated by the pink stars from Bilimogga et al. in prep) have just under an order of magnitude less gas than the field sample and an order of magnitude more than the results from stacking all Coma galaxies (magenta diamonds in \figref{fig:coma_gasfractions}). \\

    \figref{fig:coma_gasfractions} shows that there is a large spread in \hi gas fraction (\fhi) values for the Coma galaxies, however looking at the average \fhi values for all galaxies in the cluster (the pink diamonds in \figref{fig:coma_gasfractions}, which are measured from the stacked spectrum containing all individual detections and non-detections in the stellar mass bin) we note that on average, the Coma galaxies have \fhi values that are between one and two orders of magnitude lower in each stellar mass bin than the field relations and other individual measurements from less dense environments. However, while the \fhi-\mstar scaling relation compares the \hi mass to galaxies of similar stellar mass, galaxies of similar stellar mass but different morphology and size have different \hi masses \citep{Brown2015,Haynes1984a,Healy2019}. Thus, it is necessary to use a probe of the \hi content that takes morphology and size into account. 

  \subsection{H {\sc i} deficiencies}
    \label{sec:coma_def}
    
    \begin{table*}[h]
      \caption{Values used for $a$ and $b$ in \autoref{eqn:coma_mhiexp} for the different morphological types.}
      \label{tab:expmassvals}
      \centering
      \begin{tabular}{lccl}
          Sample & $a$ & $b$ & Reference \\ \hline
          All/Unknown morphology & $7.12$ & $0.88$ & \citet[][Table 5]{Haynes1984a}\\
          Early types/Ellipticals/S0s & $6.88$ & $0.89$ & \citet[][Table 5]{Haynes1984a}\\
          Late types & $7.51$ & $0.73$ & \citet[][Table 2]{Solanes1996}\\
          Irregulars & $7.45$ & $0.70$ & \citet[][Table 3]{Boselli2009}\\
      \end{tabular}
    \end{table*}
    
    The \hi deficiency parameter \citep[\defhi,][]{Haynes1984a} is a useful quantity to understand how deficient a galaxy (or sample of galaxies) is compared to a field galaxy of the same size and morphology. It is defined as:
    \begin{equation}\label{eqn:coma_hidef}
      \textsc{Def}_{\hi} = \log( \mathrm{M}_{\hi,exp} ) - \log( \mathrm{M}_{\hi,obs} ),
    \end{equation}
    where $\mathrm{M}_{\hi,exp}$ is the expected \hi mass which is calculated using an optical-\hi scaling relation that takes into account the morphology and size of the galaxy \citep{Haynes1984a,Denes2014}. \\

    The expected \hi mass ($\mathrm{M}_{\hi,exp}$) is derived from a log linear empirical relation which is calibrated using field galaxies on varying morphologies and sizes \citep{Haynes1984a},
    \begin{equation} \label{eqn:coma_mhiexp}
      \mathrm{M}_{\hi, exp} = a + b \log ( h D_{25}) ^2 - 2 \log h
    \end{equation}
    where $D_{25}$ is the diameter of the galaxy in the B-band at the \numunit{25^{th}}{\text{mag}/\text{arcsec}^2}, and $a$ and $b$ are empirical constants dependent on the galaxy's morphology (T-type). {$h$ is taken to be $0.7$ from \numunit{\mathrm{H}_0 = 70}{\mathrm{km\, s^{-1}\, Mpc^{-1}}}}. While there have been updates to the scaling relations used to estimate the \hi mass of galaxies without using T-types \citep[e.g.][]{Denes2014, Denes2016, Jones2018b}, {the deficiencies presented in this work are calculated using the T-types.} \\

    The Coma galaxies have been separated into broad morphological bins: early-type galaxies, S0s, late-type galaxies, and irregular galaxies, galaxies for which we cannot reliably determine the classification are marked as unknown. The classifications were made by eye (by three adjudicators) using colour optical images from three different surveys (SDSS, PanSTARRS and DeCALS) and $g$-band data observed using the Canada-France-Hawaii Telescope \citep{Head2015}. These classifications (where possible) have been cross-matched against the Galaxy Zoo database \citep{Lintott2011} and the Third Reference Catalogue of Bright Galaxies \citep[RC3][]{DeVaucouleurs1991} and are used to calculate the \hi deficiency. The values used in \autoref{eqn:coma_mhiexp} for the different morphological types are given in \autoref{tab:expmassvals}.  \\
    
    \begin{figure}[h]
      \centering
      \includegraphics[width=\linewidth]{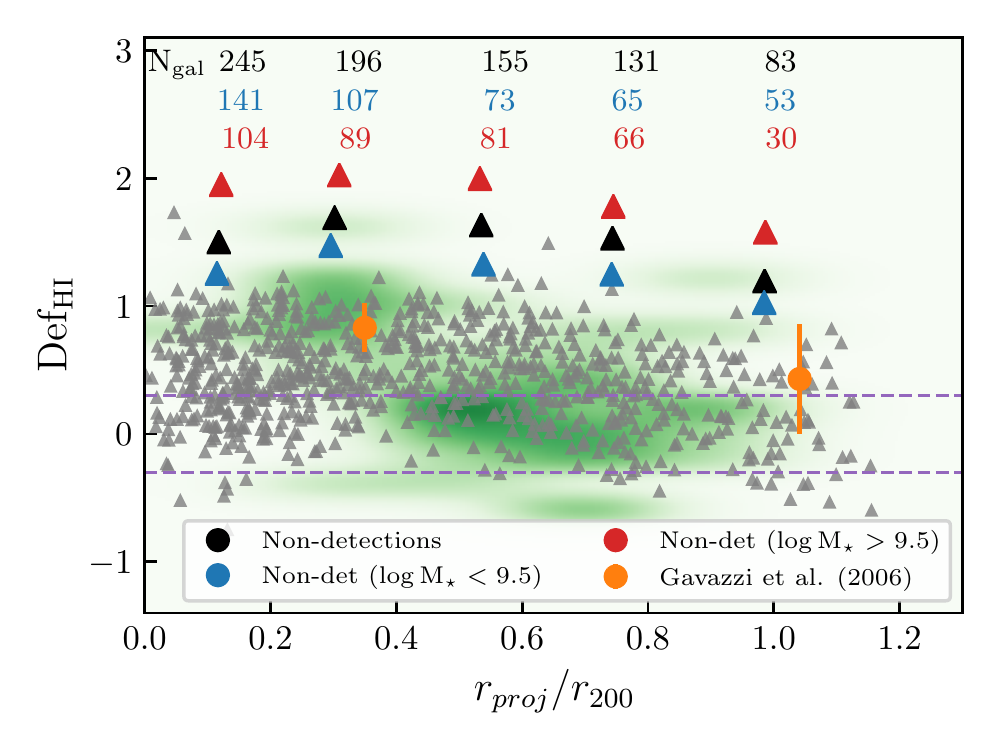}
      \caption{\hi deficiency of the Coma galaxies as a function of projected radius from the cluster centre. The individual \hi deficiency measurements for the Coma galaxies are represented by the green shaded distribution for the \hi detections and the grey triangles for the lower limits on the \hi non-detections. We also show the average \hi deficiencies for late-type Coma galaxies from \citet{Gavazzi2006} in orange. The black, blue, and red triangles represent the lower limits from stacking the non-detections in annuli in three samples: all non-detections, $\log \mstar < 9.5$ and $\log \mstar > 9.5$ respectively. {The horizontal dashed purple lines set the limit on \defhi values that indicate a galaxy is \hi-normal.} }
      \label{fig:coma_defhist}
    \end{figure}

    We calculate individual measures of the \hi deficiency for every galaxy. For the galaxies where there there is no direct \hi detection, we use the $3 \sigma$ upper limit of the measured \hi mass. This means that the corresponding \hi deficiency measurement is a lower limit and the deficiency value is expected to increase in the case of a better sensitivity spectrum. \\
    
    \citet{Denes2014} use data from the \hi Parkes All Sky Survey \citep[HIPASS,][]{Meyer2004} to map the global \hi deficiency across the southern sky. They found that regions that contained an over-density of \hi deficient galaxies correlated with known dense galaxy regions. The known galaxy clusters in the southern sky have an average \hi deficiency that lies between $0.5 <  \defhi < 2$. {For the many Coma galaxies where there is no direct \hi detection, we measure a lower limit on the \hi deficiency. These lower limits for the \defhi of the Coma galaxies (grey triangles) are predominantly between $0.5 < \defhi < 2$, which indicates that Coma is at least as deficient as the clusters covered in the \citet{Denes2014} sample}.\\
     
    For a sample of 18 clusters, \citet{Solanes2001} studied the \hi deficiency of spiral galaxies residing inside one Abell radius compared to those outside. They compare the fraction of \hi deficient spirals against different cluster properties. Coma is consistent in the total number of spiral galaxies given its x-ray luminosity, however it is an outlier in the number of \hi deficient spirals for its mass, Abell richness, and total number of spiral galaxies. \citet{Gavazzi2006} did a more detailed study on the radial pattern of \hi deficiency across the Coma Supercluster. However, like \citet{Solanes2001}, the \citet{Gavazzi2006} work only focuses on the late-type galaxies across the cluster.\\
    
    Here, we revisit the work by \citet{Gavazzi2006}, however we focus on \hi non-detections which cover all galaxy morphologies across the cluster. {To calculate the average \defhi for the \hi non-detected galaxies, we use the \hi stacking technique to stack the galaxies in annuli with increasing radius from the cluster centre, scaling each input spectrum by the galaxy's expected \hi mass (see \citet[][Eqn. 3.5]{Fabello2011} for same technique but for \hi gas fractions)}. While no clear detections in the stacked spectra (see {Figs.~\ref{fig:coma_radiiDef}.~\ref{fig:coma_radiiDeflsm}~\&~\ref{fig:coma_radiiDefhsm}} ), we do push the lower limits higher than the individual spectra (see the black triangles in \figref{fig:coma_defhist}). These results do indicate a bi-modality in the \hi content of the Coma galaxies -- we either detect the \hi, or the galaxies contain on average at least $\sim10-100$ times less \hi than their field counterparts. This suggests an extremely fast and efficient quenching mechanism. \\
    
    \citet{Oman2016} show using N-body simulations of clusters that the time for quenching low mass galaxies is longer than that of high mass galaxies. We thus separated the non-detections by stellar mass into a high stellar mass sample (\numunit{\mstar > 10^{9.5}}{\msol}{, represented by the red symbols in \figref{fig:coma_defhist}}) and a low stellar mass sample (\numunit{\mstar < 10^{9.5}}{\msol}{, represented by the blue symbols in \figref{fig:coma_defhist}}). We choose this stellar mass as it is the so-called `gas-richness' threshold \citep{Kannappan2013}: galaxies below this threshold are typically \hi gas-rich, while galaxies above this threshold are typically \hi gas-poor. {For both sub-samples, there are no detections in the stacked spectra at any radius. These results set lower limits on the average \hi deficiencies for the low and high stellar mass populations and we are therefore unable to compare to radial trends in other clusters \citep[e.g.][]{Solanes2001, Gavazzi2006} or to quenching times for different mass samples \citep{Oman2016} with these data.}

\subsection{H {\sc i} stacking in and around the cluster}
    \label{sec:stackingcluster}

    \begin{figure}[t]
      \centering
      \includegraphics[width=\linewidth]{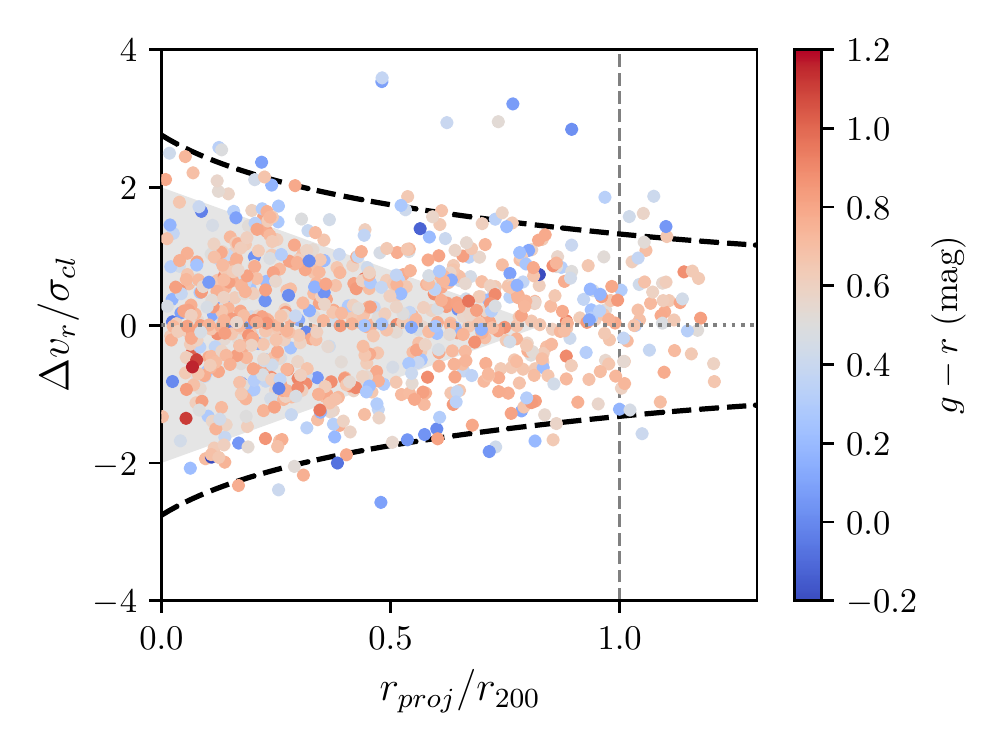}
      \caption[]{Phase-space diagram for Coma. The black dashed lines represent the escape velocity for Coma, calculated using a concentration index ($c$) of 5, \numunit{\mathrm{M_{200}} = 5.1 \times 10^{14}}{\msol} and \numunit{r_{200} = 1.8}{\text{Mpc}} \citep{Gavazzi2009}. {Galaxies inside the escape velocity `trumpet' are considered to be within the cluster, while galaxies on the outside edge of the trumpet are considered to be falling into the cluster. The light grey shaded cone represents} the virialized zone \citep{Oman2013}. The $g-r$ colour of each galaxy is given by the colour scale in colourbar. }
      \label{fig:coma_phase_colour}
    \end{figure}

    \begin{figure*}[h]
      \centering
      \includegraphics[width=\linewidth]{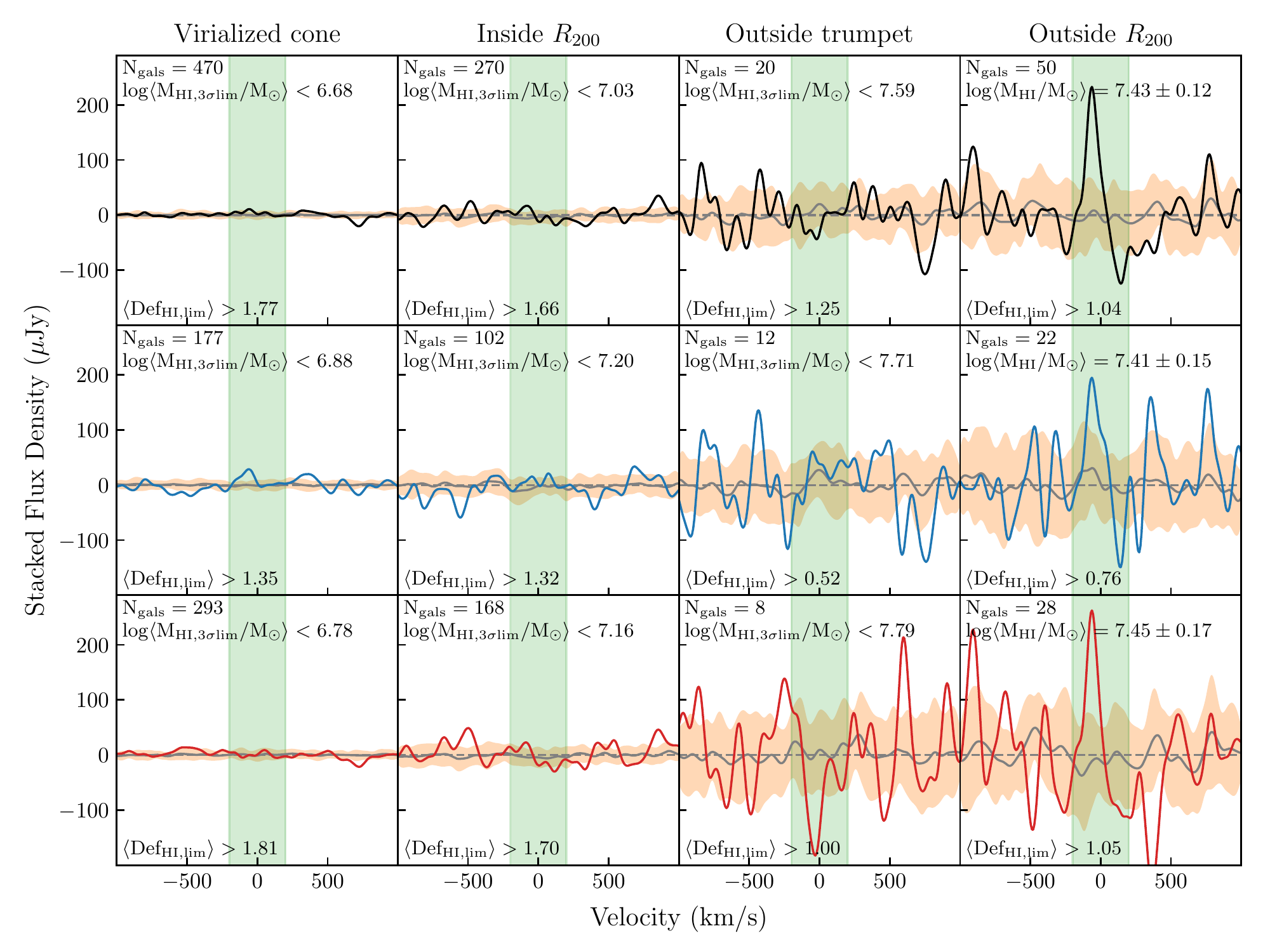}
      \caption[Stacked spectra for all, blue and red \hi non-detections in different regions of the cluster]{Stacked spectra for all (top row), blue (middle row) and red (bottom row) \hi non-detections in different regions of the cluster. The average stacked spectrum in each panel is indicated by the solid black/blue/red line. The average stacked reference spectrum is represented by the solid grey line. The green band the velocity range \numunit{cz \pm 200}{\kms} over which we would expect to see \hi emission. {Only the sample of galaxies outside the $r_{200}$ (right-most column) of the cluster appear to contain detectable \hi.}}
      \label{fig:coma_phase_stacks}
    \end{figure*}

    In this section we explore the average \hi content of the Coma galaxies. The focus here is the total galaxy sample, including galaxies without direct detections. The direct \hi detections are looked at in detail in Molnar et al. (in prep). \figref{fig:coma_phase_colour} shows the line-of-sight velocity relative to the cluster ($\Delta v_r/ \sigma_{cl}$; $\sigma_{cl}$ is the cluster velocity dispersion, \numunit{1180}{\kms}) as a function of the projected radius from the centre of the cluster ($r_{proj}/r_{200}$, where \numunit{r_{200} = 1.8}{\text{Mpc}} is the radius at \numunit{200}{\rho_{crit}}, \citealt{Gavazzi2009}), the data points are coloured by their \gmr colour. {The escape velocity for the cluster (indicated by the black dashed lines in \figref{fig:coma_phase_colour}) are determined using the prescription outlined in \citet{Jaffe2015} assuming a concentration index ($c$) of 5, and \numunit{\mathrm{M}_{200} = 5.1 \times 10^{14}}{\msol} \citep{Gavazzi2009}.} We will refer to \figref{fig:coma_phase_colour} as the phase-space diagram throughout this work. \\
    
    We separate the Coma galaxies into four samples, by the regions in the phase-space diagram in \figref{fig:coma_phase_colour}: \textit{virialised cone} (the light gray cone), \textit{inside $r_{200}$} (all galaxies with $r_{proj} < r_{200}$ and within the escape velocity `trumpet', this sample excludes the galaxies in the virialised cone), \textit{outside trumpet} (all galaxies with $r_{proj} < r_{200}$, but line of sight velocities that place them outside of the escape velocity trumpet), and \textit{outside $r_{200}$} (all galaxies with $r_{proj} > r_{200}$). The four samples are further separated by \gmr colour: galaxies with $\gmr < 0.6$ are considered to be be blue, and those with $\gmr > 0.6$ are considered to be red. For Coma, $\gmr = 0.6$ provides a good separation between the red sequence and the star-forming blue cloud. The stacked spectra for each sample (with direct \hi detections excluded) are presented in \figref{fig:coma_phase_stacks}. It should be acknowledged that the four samples (virialized zone, inside $r_{200}$, outside trumpet and outside $r_{200}$) will contain interlopers due to projection effects especially along the line-of-sight velocity. These projection affects are most likely to affect the three samples contained within the $r_{200}$ radius of the cluster. \\
    
    In \figref{fig:coma_phase_stacks} we compare the average \hi content of the galaxies for which there are no direct \hi detections for the three different regions of the cluster. We also separate the galaxies by colour, the resulting average spectra for the blue and red galaxies in each region are shown in the middle and bottom rows of \figref{fig:coma_phase_stacks}. The stacked spectra are smoothed after stacking to a velocity resolution of \numunit{40}{\kms} to improve the signal-to-noise ratio (S/N). We also stack the 25 reference spectra (see \secref{sec:coma_histackingmethod} using the same scheme as the target spectra. From the 25 stacked reference spectra, we create an average reference spectrum which is represented by the grey spectra in \figref{fig:coma_phase_stacks}. Also shown in \figref{fig:coma_phase_stacks} is the variance of the 25 stacked reference spectra, this is represented by the orange band around the grey line. The comparison between the average reference spectrum, where we do not expect to see a signal, and the stacked target spectrum provides confidence that any signal visible in the target spectrum is real. We consider stacked spectra with a signal in the region \numunit{v = 0\pm 200}{\kms} (represented by the vertical green band in \figref{fig:coma_phase_stacks}) that has $\mathrm{S/N} = 3.5$ to be marginally detected. We feel confident using a $\mathrm{S/N} = 3.5$ since there is no corresponding signal in the reference spectrum.\\
    
    Only the sample of galaxies outside $r_{200}$ shows a possible detection in all three colour samples{, despite both the red and blue samples having roughly the same average \hi mass (\ave{\mhi}), it should be noted that the lower limits on the average \hi deficiency (\ave{\defhi}), which is shown in the lower left corner of each panel, show the red galaxies to be more \hi deficient than the blue galaxies. Despite both the blue and red samples showing lower limits for the \hi deficiency, the two samples have similar number of galaxies meaning that the stacked spectra have similar sensitivities making their limits comparable. The difference between the \defhi for the two samples is not unexpected as red galaxies are well known for being more gas poor than blue galaxies. Unlike what is found by \citet{Jaffe2016} for Abell 963, none of the samples show any possible detections within the $r_{200}$, however they also do not detect a signal for their red galaxies outside the $r_{200}$, which we do.} \\
    
    Simulations have shown that quenching begins after some delay time once a galaxy has crossed $2.5\,v_{vir}$ (for reference, $r_{200} \sim 0.73 r_{vir}$, \citealt{Oman2016}) with higher mass galaxies affected slightly quicker than lower mass galaxies \citep{Oman2016}, however once quenching has begun, the time taken for the galaxies to transition from a field-like galaxy to one fully processed by the cluster is very short. Indeed, less than 10 of our galaxies outside $r_{200}$ have measurable star formation rates, and most of those are below the star forming main sequence (see panel b of \figref{fig:coma_smass_sfr} for galaxies with $7 < \log \mstar/\msol < 9.5$).\\
    
    \citet{Gavazzi2006} show on average Coma galaxies at \numunit{r > 5}{\text{Mpc}}, which is beyond $2\,r_{vir}$, are \hi normal. We have shown in the previous section, on average, that our Coma galaxies are (at a lower limit) at least 10--100 times more \hi deficient than their field counterparts. If we assume that galaxies that are accreted onto the cluster are \hi normal and on the star forming main sequence when they cross $2.5\,r_{vir}$, the processes driving the removal of \hi in galaxies as they fall into the cluster must be very quick and have a strong influence beyond $r_{vir}$. In \secref{sec:coma_locaglobal} we also compare the average \hi content between the Coma galaxies in groups/substructure to those that are not associated with any groups.
        
\section{Finding substructure in Coma}
      \label{sec:coma_substructure}    
      Substructure, defined as structure kinematically distinct from the parent halo, is a natural consequence of a hierarchical universe \citep{Hou2012}. As discussed in the introduction, there are many different methods that can identify substructure within a galaxy cluster \citep[e.g.][]{Adami2005,Dressler1980,Jaffe2013,Neumann2003}. We are interested in how the average \hi gas content may change in groups/substructure with different morphologies or possible infall times across the cluster. \\
      
      To identify these groups/substructure, we choose to use the Dressler-Shectman test to identify groups that are kinematically distinct from the cluster. \citet{Dressler1988} applied their test for kinematic deviations to look for substructure in a number of Abell clusters (including Coma). Their conclusion regarding Coma was that there was no substructure within the cluster. Since \citet{Dressler1980}, there have been many targeted and blind spectroscopic surveys of the Coma cluster and subsequent studies of Coma have found significant substructure \citep[e.g.][]{Adami2005,Colless1996}. In order to detect substructure, one needs $\sim 1000$ redshifts in a \numunit{\sim 1}{\text{Mpc}} region \citep{Adami2005}. We apply the DS test, given by:
      \begin{equation}\label{eqn:dstest}
        \delta_i^2 = \left( \frac{N_{nn} + 1}{\sigma_{cl}^2} \right) \left[ ( \bar{v}_{local}^i - \bar{v}_{cl})^2 + (\sigma_{local}^2 - \sigma_{cl}^2) \right],
      \end{equation}
      to our redshift catalogue. We use nearest neighbour ($N_{nn} $) values in increments of five from five to thirty with a velocity dispersion ($\sigma_{cl}$) of \numunit{1180}{\kms} and a cluster velocity ($v_{cl}$) of \numunit{6925}{\kms}. \figref{fig:coma_DStest} shows the results of the DS Test for $N_{nn} = 25$. Using similar plots for all the $N_{nn} $ values, and two independent judges, we identify groups where there are large overlapping circles (which represent large $\delta_i$ values) of similar colour representing similar velocity values. Using the $N_{nn} = 25$ as a reference, galaxies were assigned to groups using a combination of the different $N_{nn} $ runs.\\
    
\begin{figure*}[h]
        \centering
        \includegraphics[width=\linewidth]{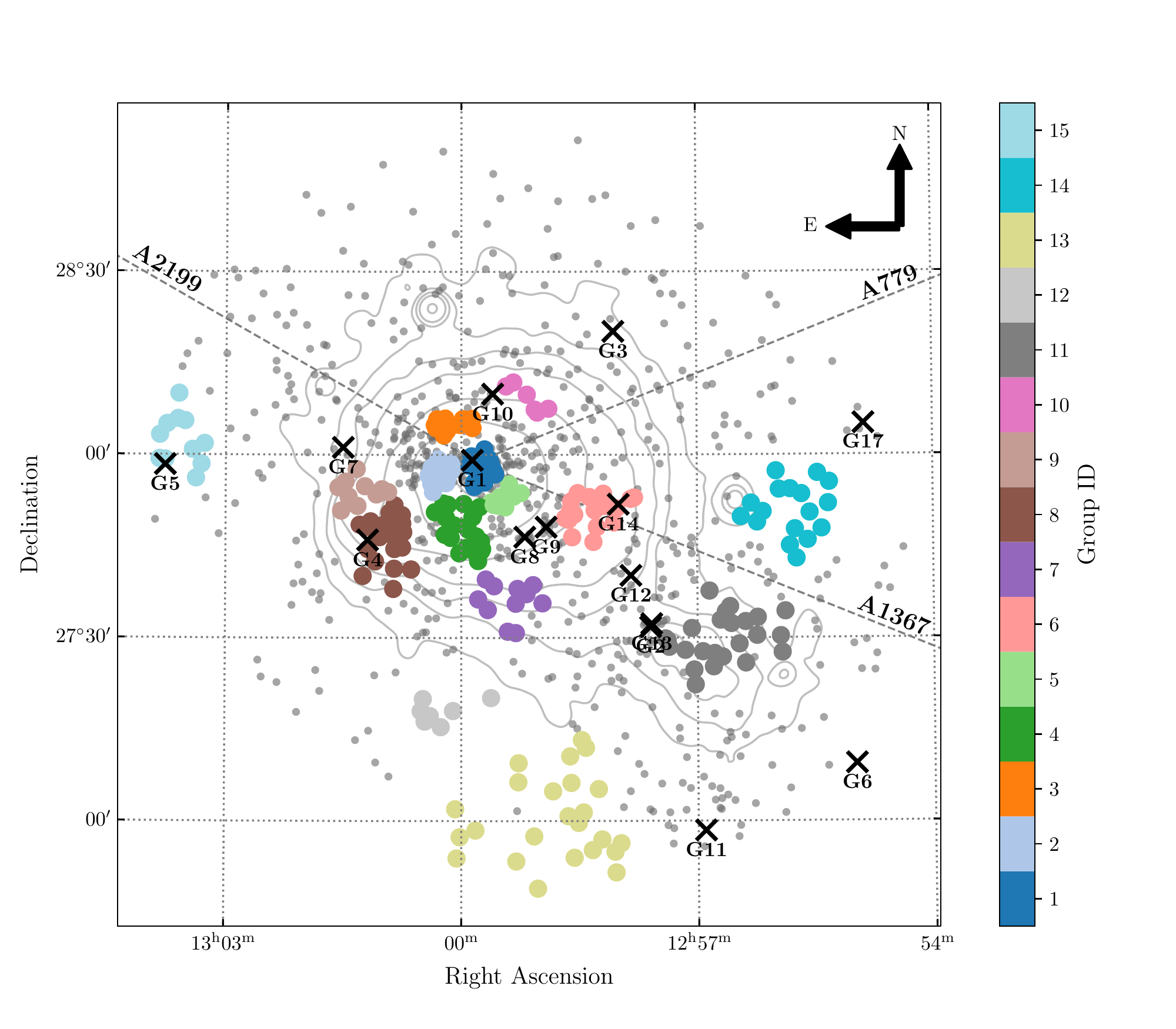}
        \caption[Coma substructure]{\textbf{Substructure in the WCS footprint. The 15 groups we identified are shown in the different colours}. The background grey contours represent the x-ray emission measured by ROSAT in \numunit{0.4-2.4}{\text{keV}}. The dashed grey lines indicate the direction to nearby clusters connected to Coma. The groups identified by \citet{Adami2005} are labelled and marked by the black crosses.}
        \label{fig:coma_substructure}
      \end{figure*}

      Within the WCS \hi data footprint, we find 15 distinct groups. These groups stand out across the six iterations of the DS test, and are highlighted by the different colours in \figref{fig:coma_substructure}. The properties of the 15 groups are given in \tabref{tab:coma_groups}. Despite the difference in the method of identifying substructure, we find some overlap between our groups and those identified by \citet{Adami2005} who use a hierarchical clustering algorithm \citep[see][]{Serna1996} which uses a measure of the binding energy to find groups of galaxies that are bound in substructure. Of the 17 groups \citet{Adami2005} found, 14 are within the WCS footprint and seven correspond to groups that we find using the DS Test. The groups overlapping between our work and \citet{Adami2005} are S1, S2, S6 and S11 (these are combined into one group at the core of the cluster), S8, S9, and S15.

      \begin{table*}
        \caption{Details of the 15 groups found within the Coma cluster using the DS test.}
        \label{tab:coma_groups}
        \centering
        \footnotesize{
        \begin{tabular}{ccccccccc}
        Group ID     & Members\tablefootmark{a} &  $\ave{v}$ & $\sigma_{group}$ & E:S0:L:I:U\tablefootmark{b} & $ \log \ave{\mhi/\msol}$\tablefootmark{c}    &  $\log \ave{\mstar/\msol}$ & $\ave{\defhi}$\tablefootmark{d} & NGC members   \\ 
           &   &  (\kms) & (\kms) &  & (\msol) &  (\msol) &  &  \\\hline
         S1 & 27 & 7699 & 924 & 8:10:0:5:4 & < 7.20 & 10.84 & > 0.72 & NGC4874 \\ 
         S2 & 18 & 6113 & 932 & 6:6:0:5:1 & < 7.43 & 10.80 & > 1.00 & NGC4889, NGC4894 \\ 
         S3 & 14 & 7626 & 358 & 7:3:0:4:0 & < 7.34 & 9.87 & > 0.67 &  \\ 
         S4 & 22 & 7981 & 899 & 6:7:1:5:3 & < 7.28 & 10.40 & > 1.08 &  \\ 
         S5 & 10 & 5868 & 843 & 5:5:0:0:0 & < 7.54 & 10.25 & > 1.57 & NGC4869 \\ 
         S6 & 17 & 6095 & 884 & 9:7:0:1:0 & < 7.39 & 9.82 & > 1.57 &  \\ 
         S7 & 11 (2) & 6137 & 1031 & 4:3:2:2:0 & 7.37 & 9.83 & > 1.40 &  \\ 
         S8 & 22 (1) & 7386 & 1238 & 11:5:1:3:2 & 7.76 & 10.15 & > 1.32 & NGC4919, NGC4911 \\ 
         S9 & 11 (1) & 6498 & 942 & 7:3:1:0:0 & 8.22 & 10.64 & 1.13 & NGC4923, NGC4921 \\ 
         S10 & 6 (1) & 8355 & 895 & 1:4:1:0:0 & < 7.71 & 10.62 & 0.81 & NGC4860, NGC4858 \\ 
         S11 & 24 & 7621 & 462 & 12:6:0:2:4 & < 7.36 & 10.52 & > 1.12 & NGC4839 \\ 
         S12 & 7 & 7584 & 460 & 4:1:0:2:0 & < 7.69 & 9.68 & > 1.23 &  \\ 
         S13 & 24 (3) & 7351 & 691 & 13:5:1:5:0 & 8.13 & 10.30 & 0.74 & NGC4859, NGC4892 \\ 
         S14 & 17 & 6898 & 522 & 10:7:0:0:0 & < 7.59 & 10.25 & > 1.45 &  \\ 
         S15 & 11 & 6436 & 603 & 7:3:1:0:0 & < 7.93 & 10.22 & > 1.16 & NGC4943, NGC4934 \\ \hline
        \end{tabular}
        }
        \tablefoot{
        \tablefoottext{a}{Numbers in brackets are the number of individual direct detections in the group.}
        \tablefoottext{b}{Number of Elliptical/Early type galaxies (E), S0/Lenticulars (S0), Late-types (L), Irregular galaxies (I), \& galaxies of Unknown morphology (U).}
        \tablefoottext{c}{Average \mhi obtained from stacking, where there is no detection in the stacked spectrum, a $3\sigma$ upper limit is used.}
        \tablefoottext{d}{Average \defhi obtained from stacking, where there is no detection in the stacked spectrum, a $1\sigma$ lower limit is used.}
        }
        \end{table*}

    \section{Identified substructures}
      \label{sec:coma_subs_discuss}

      \subsection{Core substructures}
        \label{sec:coma_coresubs}

          In the core of the cluster (see \figref{fig:coma_substructure}), we identify two groups (groups S1 and S2) around the two cD galaxies (NGC 4874 and NGC 4889) and another two groups surrounding the two central groups: S3 and S5. There is much discussion in the literature surrounding the centre of the Coma cluster. A commonly accepted scenario is that the two cD galaxies will eventually merge \citep[e.g.][]{Colless1996, Adami2005}.\\
          
          Using the hierarchical clustering method to developed by \citet{Serna1996} to identify substructure, \citet{Adami2005} find one group (G1) in the central region of the cluster. They later note that with x-ray analysis there are two distinct peaks in the x-ray corresponding to the two galaxies, and postulate that these two groups are in the process of colliding. S1, the larger of the two groups, with 27 members has a mean velocity of \numunit{7699}{\kms}, and S2 (with 18 members) has a mean velocity of \numunit{6113}{\kms}. Both the groups have velocity dispersions that are of the order of \numunit{\sim 900}{\kms} (see \tabref{tab:coma_groups} for more details).\\

\begin{figure*}[t]
            \centering
            \includegraphics[width=\linewidth]{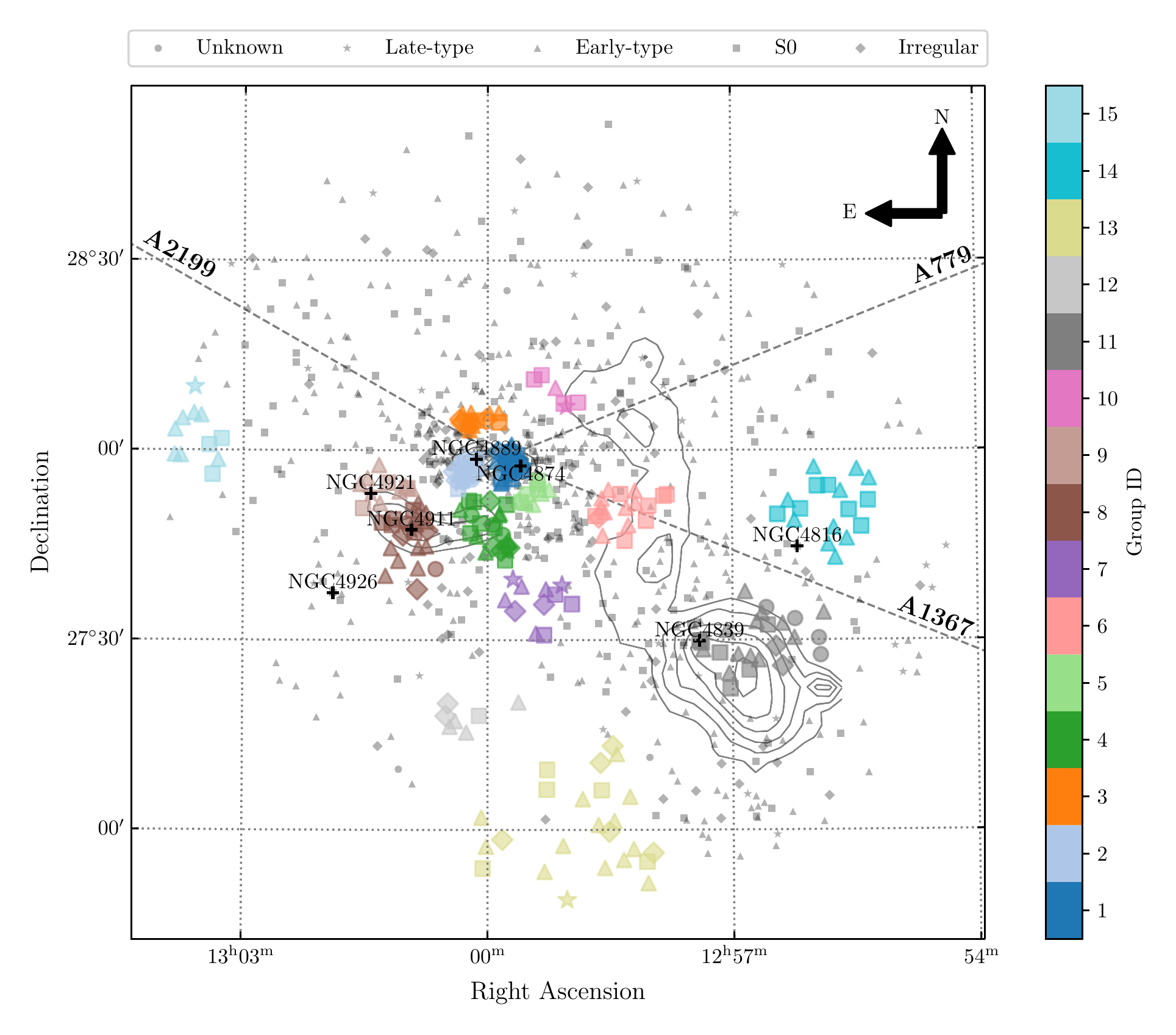}
            \caption[Coma groups with x-ray excess]{The 15 identified groups within the WCS footprint. The contours (taken from \citet[][Fig. 4]{Adami2005}) represent the x-ray excess found by \citet{Neumann2003}.}
            \label{fig:coma_subxray}
        \end{figure*}

          S3 contains 14 members and is located immediately north of S1 and S2 (see \figref{fig:coma_subxray}). S5, the smallest group containing only six galaxies, is located to the south-west of S1. Both S3 and S5 groups are dominated by early-type and S0 galaxies. None of these groups (S1, S2, S3 and S5) show a detection of \hi in their stacked spectra (see \figref{fig:coma_group_stacks1}), however given the gas removal processes at play in cluster centres, this is unsurprising. The measured upper limits on the \hi mass and lower limits on the \hi deficiency for the groups can be found in \tabref{tab:coma_groups}, and the corresponding stacked spectra in \appendref{sec:coma_stackedspectra}.

      \subsection{Groups coincident with excess x-ray emission}
        \label{sec:coma_xraysubs}

        \subsubsection*{South west structure}
        Using x-ray observations by XMM-Newton, \citet{Neumann2003} study the dynamical structure within Coma. They fit an elliptical beta model to the x-ray emission where any residuals from this fit may indicate structure within the cluster. Fig. 2 in \citet{Neumann2003} shows that there is x-ray emission not explained by the beta model. \citet{Neumann2003} also find a number of bright galaxies coincident with the peaks in this excess emission. \citet{Adami2005} compare the groups that they find to the excess x-ray emission maps from \citet{Neumann2003}, and find a number of overlaps. We use the excess x-ray contours from \citet[][Fig. 4]{Adami2005} to compare to the location of our groups. \figref{fig:coma_subxray} shows the isocontours residual x-ray emission along with the location of our groups. From \figref{fig:coma_subxray}, one can see that groups S4, S6, S8, S9, S10 (this group is slightly offset from the x-ray), and S11 are coincident with x-ray emission. \\

        Of the groups we identify that are coincident with x-ray emission, S11 is perhaps the most well studied \citep[e.g.][]{Colless1996, Neumann2001, Akamatsu2013,Burns1994, Lyskova2018,Beijersbergen2003}. S11 is located to the south-west of the cluster core and is home to NGC 4839. Previously, \citet{Beijersbergen2003} found a group of 18 galaxies within $0.15^\circ$ of NGC 4839, and using data from their blind WSRT \hi survey of the cluster, calculated for the group, an average \hi mass (\mhi) $3\sigma$ upper limit to be \numunit{\ave{\mhi} < 7.55\times 10^8}{\msol}. Within the same radius, we find 26 galaxies, however using the DS test we find 36 galaxies associated with this group. Our average $3\sigma$ mass limit is \numunit{\ave{\mhi} < 2.29\times 10^7}{\msol}. Many groups have studied this collection of galaxies in an effort to understand if it is still on first infall \citep[e.g.][]{Colless1996, Neumann2001, Akamatsu2013}, or has already passed through the centre of the cluster \citep[e.g.][]{Burns1994, Lyskova2018}.\\
        
        \citet{Lyskova2018} show using smooth particle hydrodynamical simulations that a post-merger scenario where the group has already passed around the centre of the cluster explains the observed x-ray emission better than the pre-merger scenario where this group is on first infall. In the post-merger scenario, \citet{Lyskova2018} postulate that S11 entered the cluster from a filament on the north-east side of the cluster. \citet{Malavasi2019} use the DisPerSE algorithm to identify filaments connecting to Coma, and while they identify three such filaments (including one joining to the north-east of the cluster), they do not identify the filament known as the ``Great Wall'' to the south-west of Coma which connects the cluster to Abell 1367. It is the ``Great wall'' filament that S11 is thought to be accreted from \citep{Colless1996, Akamatsu2013}. From our results, we cannot conclude whether these groups are on first infall, or have already passed through the cluster. However, our results do indicate that this group is extremely \hi deficient and contains an evolved population of galaxies which supports the post-merger scenario, but also first infall under the assumption that this group has been pre-processed. Recent work by \citet{Salerno2020} has shown that filaments can be very effective at quenching star formation. Nevertheless, the question whether the NGC 4839 (S11) group is on first infall or has already passed through the cluster, remains open.\\

        \subsubsection*{South eastern structure}
        \citet{Neumann2003} identify x-ray substructure to the south east of the core of the cluster (see S4, S8, and S9 in \figref{fig:coma_subxray}). \citet{Neumann2003} attribute this x-ray overdensity to emission from the intragroup medium of one group of galaxies containing NGC 4911 and NGC 4921. They attribute the bending in the structure due to projection effects by assuming that the direction of the structure motion is to the east with NGC 4921 at the head. They also conclude that NGC 4911 and NGC 4921 are gravitationally bound due to similar velocities. We do not find these two galaxies to be part of the same group, which can largely be attributed to the differences in their velocities (\numunit{\sim 2000}{\kms}). The velocity we use for NGC 4926 is taken from SDSS DR13. The SDSS DR13 measurement is in good agreement with other optical and \hi redshift measurements of this galaxy \citep[e.g.][]{Dressler1988a,DeVaucouleurs1991,Zabludoff1993,Haynes1997,Bravo-Alfaro2000}, including the \hi measurement from the WCS. \citet{Adami2005} also find NGC 4911 and NGC 4926 to be part of different groups and note that the velocity quoted by \citet{Neumann2003} is incorrect. \\
      
        We identify three groups (S4, S8, and S9) coincident with this x-ray substructure, S8 and S9 have been previously identified by \citet{Adami2005} as G4 and G7 respectively. S8 contains NGC 4911 which appears to spatially coincident with the peak of the x-ray emission \citep{Adami2005, Neumann2003}. If the x-ray is indeed associated with these groups, the fact that it is still detectable implies that the groups have not yet passed through the cluster \citep{Adami2005}. However the x-ray emission shows no sign of shock heated gas which is a signature of radial infall, thus \citet{Adami2005} argue that it is more likely that these groups are on a spiral infall trajectory. Of the three groups, S4 and S8 have mean velocities that are higher than the cluster velocity,{ S4 and S8 also contain a number of late-type and irregular galaxies (in addition to the E and S0 galaxies) whereas S9 is dominated by early-type galaxies (E or S0) and has only one late-type galaxy (see \tabref{tab:coma_groups})}. Both S4 and S8 are on average extremely \hi deficient ($\ave{\defhi} > 1$, see \tabref{tab:coma_groups}), thus it is possible that the galaxies have been stripped of their \hi, a process that can occur extremely rapidly after crossing into the cluster (\numunit{\sim 2.5}{r_{vir}}), but the presence of the bluer galaxies may indicate that they have yet to transition across the `green valley' \citep{Oman2016}. S9 is dominated by an older population of galaxies, and has a mean velocity of \numunit{6498}{\kms} which is similar to that cluster velocity which could indicate that it is moving in the plane of the sky. The presence of one \hi detection in this group would imply that it could also be a recent infall, however given the difference in mean velocities from S4 and S8, it is unlikely to have followed the same trajectory. 
        
        \subsubsection*{Western filament}
        The last major x-ray residual is the western filament which is aligned north-south, extending roughly a megaparsec north from NGC 4839. \citet{Neumann2003} discuss possible origins of this structure: the change in x-ray temperature between the western filament and the centre of the cluster suggests that it could be due to heating from compression or shock waves during the infall of a structure onto the core of the cluster. They also discuss the possibility that the two x-ray maxima in the filament are due to a galaxy group that was broken up on infall; they discount the possibility of two groups falling in to the cluster at the same time. However, \citet{Neumann2003} do not find any galaxy over-densities associated with the filament.\\
        
        \citet{Adami2005} find two groups coincident with the western filament: G12 and G14 (see \figref{fig:coma_substructure}). We also find two groups coincident with the filament: S6 and S10 (see \figref{fig:coma_subxray}). S6, which is located about half way up the filament, is coincident with G14 found by \citet{Adami2005}. Our other group that is coincident with the filament, S10, is located at the northern most part of the structure. \\
        
        Both S6 and S10 lie on lines that connect Coma to other clusters: S6 on the line connecting to Abell 1367, and S10 on the line connecting to Abell 779 (see \figref{fig:coma_subxray}). Both of these groups are dominated by early-type and S0 galaxies, however, S10 contains one direct \hi detection which dominates the stacked spectrum (see \figref{fig:coma_group_stacks1}). S10 has a higher average velocity (\numunit{cz = 8355}{\kms}) than Coma which could indicate that it is falling in from the filament connected to Abell 779. The average velocity of S6 is similar to Coma, which like S9, could either indicate that the group is moving slowly or in the plane of the sky.

    \subsection{Groups near the outskirts}
      \label{sec:coma_outgroups}

        There are four groups which can be considered on the outskirts: S12, S13, S14 and S15. Of these four groups, only S15 had been previously identified by \citet{Adami2005} as G5. \citet{Adami2005} identify five other groups on the outskirts of the cluster which we do not identify in this work, it should be noted that two of those groups are outside our \hi dataset and the other two are on the edge. Despite using a galaxy catalogue that covered a larger area than our \hi dataset, we only identified groups for which we had \hi data. The one group contained within our \hi dataset by \citet{Adami2005} that we do not find, G11, is located around $\alpha = 12^h57^m \, \delta=27^\circ00'$. It is clear from \figref{fig:coma_DStest} which shows the results of the DS test, that there are no overlapping large circles of similar colour. Given that our redshift catalogue has more redshifts than \citet{Adami2005}, in particular in that region due to our observing campaign with WIYN (see \secref{sec:coma_wiynredshifts}), it is possible that with the smaller number of redshifts, G11 was artificially highlighted.\\
        
        S13 has a diverse population of galaxies and contains the highest number of direct \hi detections, while the other three groups (S12, S14 and S15) are dominated by older early-type or S0 galaxies with no direct \hi detections. S14 and S15 both lie on lines to other clusters (Abell 2199 and Abell 1367 respectively) and have similar velocities to the cluster indicating slow movement or a velocity vector in the plane of the sky. If S14 is being accreted onto Coma, then the average \hi deficiency of the group and the evolved galaxy morphologies could indicate that this group underwent advanced evolutionary processes prior to being accreted. The high average velocity of S13 (\numunit{cz = 7351}{\kms}) could indicate that it is falling into Coma from the foreground, this is supported by the fact that this group is not in line with any of the known filaments connecting to Coma. Given that S13 also contains so many direct detections, this also may suggest that it is being accreted from a low density environment.  

  \subsection{\hi in substructure vs. the cluster}
    \label{sec:coma_locaglobal}
    
    \begin{figure*}[h]
        \centering
        \includegraphics[width=\linewidth]{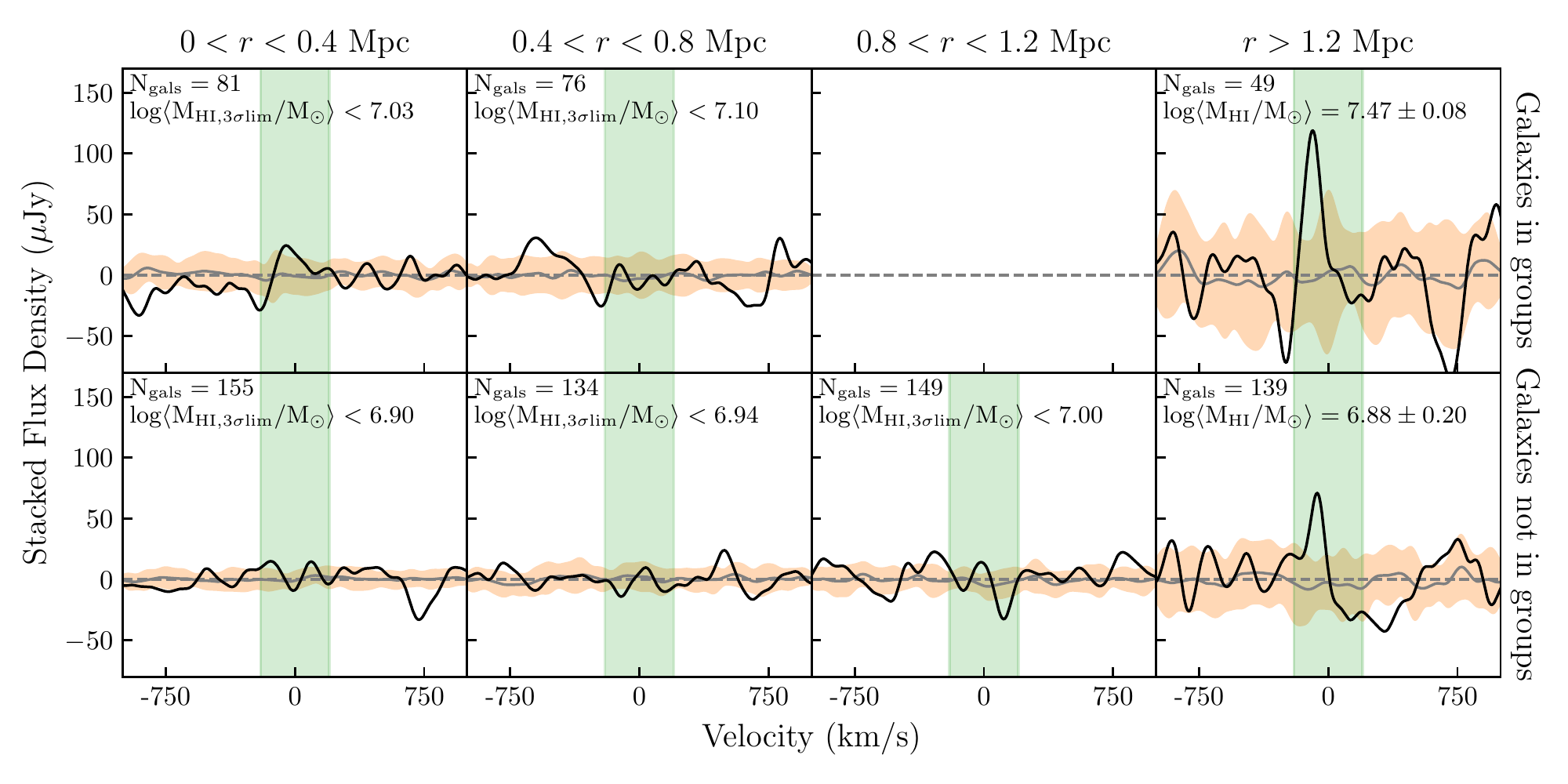}
        \caption{Stacked spectra for galaxies in groups (top row), and galaxies not in groups (bottom row) separated into different annuli from the cluster centre. The average stacked spectrum in each panel is indicated by the solid black line. The average stacked reference spectrum is represented by the solid grey line, with the variance of 25 stacked reference spectra surround the average reference spectrum in orange. The green band the velocity range \numunit{cz \pm 200}{\kms} over which we would expect to see \hi emission.}
        \label{fig:coma_rings_stacks}
    \end{figure*}
    
    \begin{figure*}[h]
        \centering
        \includegraphics[width=\linewidth]{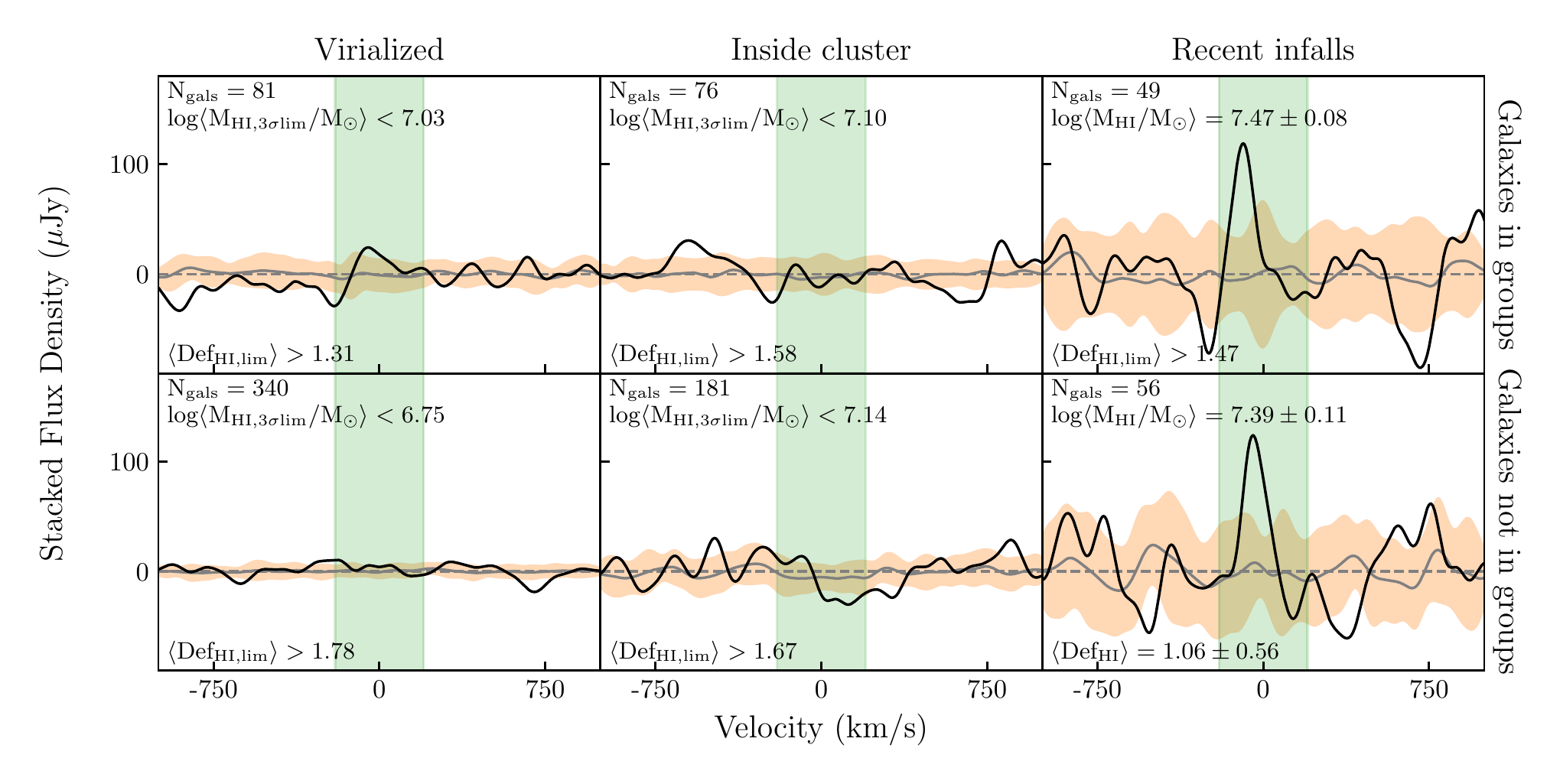}
        \caption{Stacked spectra for galaxies in groups (top row), and galaxies not in groups (bottom row) separated into different regions in the cluster. The average stacked spectrum in each panel is indicated by the solid black line. The average stacked reference spectrum is represented by the solid grey line, with the variance of 25 stacked reference spectra surround the average reference spectrum in orange. The green band the velocity range \numunit{cz \pm 200}{\kms} over which we would expect to see \hi emission.}
        \label{fig:coma_}
        \label{fig:coma_groupphase}
    \end{figure*}    
    
     We explore the \hi content of the Coma galaxies in substructure compared to Coma galaxies not in substructure. We separate the Coma galaxies into \numunit{0.4}{\text{Mpc}} annuli increasing from the cluster centre. The 15 substructures are also separated into the different annuli: S1-S5 (\numunit{0 - 0.4}{\text{Mpc}}), S6-S11 (\numunit{0.4 - 0.8}{\text{Mpc}}), and S12-S15 (\numunit{>1.2}{\text{Mpc}}). \figref{fig:coma_rings_stacks} shows the stacked spectra for galaxies in groups in the different annuli (top row) as well as galaxies not in groups (bottom row). There are no possible detections in any of the stacked spectra in \figref{fig:coma_rings_stacks} with the exception of the galaxies with \numunit{r > 1.2}{\text{Mpc}}. This is perhaps not surprising as we did not take into account the velocity of the galaxies when separating them into the different annuli. It is possible that there are infalling galaxies in the two innermost annuli; however, this inner region is typically dominated by the oldest members of the cluster.  \\
     
    In order to explore the difference in \hi content between galaxies not associated with any substructure and those that reside in substructure in the different regions of the cluster, we separate the two groups of galaxies into three bins loosely based on the regions of the phase-space diagram (\figref{fig:coma_phase_colour}). The galaxies and substructures that reside predominantly in the virialized cone are assigned to the `virialized bin'; galaxies/substructures inside $r_{200}$ are assigned to the `inside cluster' bin, and the remaining galaxies/substructures which predominantly have $r_{proj} > r_{200}$ are assigned to the `recent infalls'. The resulting stacked profiles for the six samples are presented in \figref{fig:coma_groupphase}. The top panel of \figref{fig:coma_groupphase} shows the stacked spectra from the galaxies in substructure and the bottom panel shows the stacked spectra from the galaxies not associated with any substructure. As with all the previous stacking, we only examine the galaxies for which there are no direct \hi detections.\\
     
    We note detections in the stacked spectra for the 'recent infalls' samples, both for galaxies in groups and galaxies not associated with any substructure (right-most panels of \figref{fig:coma_groupphase}). It is not unexpected that only the 'recent infalls' show detections in the stacked spectrum. {The average \hi deficiency measurement for each sample is given in the lower left corner of each panel in \figref{fig:coma_groupphase}.} It is interesting to note that while the two samples (galaxies within substructure and galaxies not in substructure) have similar average \hi masses, they differ in average \hi deficiencies. Galaxies in groups have a lower limit on the average \hi deficiency of $\ave{\defhi} > 1.47$, while the non-group galaxies have an average \hi deficiency of $\ave{\defhi} = 1.05 \pm 0.56$. The galaxy size, stellar mass and colour distribution for both samples is similar, and while the \ave{\defhi} measurements indicate that both samples are \hi deficient, the fact that we can only measure a lower limit on the average \hi deficiency for the group galaxies suggests that they are more deficient than the non-group galaxies for which we can measure an average \hi deficiency. This could indicate that the gas removal process is more efficient for the group galaxies.

\section{Summary}

    In this work we have aimed to study the average \hi content and morphologies of galaxies in the substructure of the Coma cluster. We have collated a large catalogue of redshifts for the Coma cluster, including 59 new redshifts that we observed with the multi-object spectrometer, Hydra, at WIYN. Using the newly compiled catalogue, we applied the DS test to find substructure within the cluster. Fifteen substructures were found, of which three had not been previously identified. \\
    
    The Westerbork Coma Survey (WCS) provided high resolution and high sensitivity \hi data that covered both the cluster core and the NGC 4839 group to the south-west. However, with only 39 direct \hi detections out of the 850 galaxies that lie within the WCS footprint, we needed an alternate method to study the \hi content of the galaxies in the different groups. The \hi stacking technique enabled us to probe the average \hi content of samples of galaxies. \\
    
    Using the \hi stacking technique, we are able to probe $\sim$1--2 orders of magnitude lower in \hi mass. Despite this improvement in sensitivity, for many of the stacked samples there is no detection in the stacked spectrum. The \hi gas fraction to stellar mass relation as well as a study of the \hi deficiency indicates that Coma galaxies are at least 50 times more \hi deficient than their field counterparts. The bi-modality in the \hi content where galaxies either have detectable \hi content or contain orders of magnitude less \hi than expected supports the theories that favour extremely rapid quenching and gas removal mechanisms. \\
    
    Given the effectiveness of gas removal processes increases closer to the centre of the cluster, we expect to see the average \hi content increase with distance from the cluster centre. For most samples of galaxies inside the cluster $r_{200}$, there is no \hi detectable in the stacked spectra, however we do detect \hi in both blue and red galaxies outside the $r_{200}$. When we separate the outer galaxies into those that are in groups versus those not in groups, both samples show detections in the `recent infall' galaxies (see \figref{fig:coma_groupphase}). The average \hi masses for the two samples are similar, however {we note that the average \hi deficiency for the group galaxies is not detected at a lower limit that is higher than the measured average \hi deficiency of the non-group galaxies, implying that the group galaxies are more \hi deficient.}. \\

\begin{acknowledgements}
    We thank the anonymous referee for their comments that have improved this work. This project has received funding from the European Research Council (ERC) under the European Union’s Horizon 2020 research and innovation programme (grant agreement no. 679627; project name FORNAX). JH acknowledges research funding from the South African Radio Astronomy Observatory. We thank R Healy for useful comments that improved the readability. Based on observations at Kitt Peak National Observatory, National Optical Astronomy Observatory (NOAO Prop. ID: 2017A-0356; PI: J. Healy), which is operated by the Association of Universities for Research in Astronomy (AURA) under a cooperative agreement with the National Science Foundation. JMvdH acknowledges support from the European Research Council under the European Union’s Seventh Framework Programme (FP/2007-2013)/ERC Grant Agreement no. 291531 (HIStoryNU).\\
    Funding for the Sloan Digital Sky Survey IV has been provided by the Alfred P. Sloan Foundation, the U.S. Department of Energy Office of Science, and the Participating Institutions. SDSS acknowledges support and resources from the Center for High-Performance Computing at the University of Utah. The SDSS web site is \url{www.sdss.org}.\\
    SDSS is managed by the Astrophysical Research Consortium for the Participating Institutions of the SDSS Collaboration including the Brazilian Participation Group, the Carnegie Institution for Science, Carnegie Mellon University, the Chilean Participation Group, the French Participation Group, Harvard-Smithsonian Center for Astrophysics, Instituto de Astrofísica de Canarias, The Johns Hopkins University, Kavli Institute for the Physics and Mathematics of the Universe (IPMU) / University of Tokyo, the Korean Participation Group, Lawrence Berkeley National Laboratory, Leibniz Institut für Astrophysik Potsdam (AIP), Max-Planck-Institut für Astronomie (MPIA Heidelberg), Max-Planck-Institut für Astrophysik (MPA Garching), Max-Planck-Institut für Extraterrestrische Physik (MPE), National Astronomical Observatories of China, New Mexico State University, New York University, University of Notre Dame, Observatório Nacional / MCTI, The Ohio State University, Pennsylvania State University, Shanghai Astronomical Observatory, United Kingdom Participation Group, Universidad Nacional Autónoma de México, University of Arizona, University of Colorado Boulder, University of Oxford, University of Portsmouth, University of Utah, University of Virginia, University of Washington, University of Wisconsin, Vanderbilt University, and Yale University.\\
    This publication makes use of data products from the Wide-field Infrared Survey Explorer, which is a joint project of the University of California, Los Angeles, and the Jet Propulsion Laboratory/California Institute of Technology, funded by the National Aeronautics and Space Administration.
\end{acknowledgements}

%
%
\bibliographystyle{aa}
\bibliography{references}

\begin{appendix}
    \section{Miscellaneous Figures and Tables}

\begin{table}[h]
    \centering
    \caption{Number of redshifts obtained from the different literature sources discussed in \secref{sec:coma_litredshifts}.}
    \label{tab:litredshiftlist}
    \begin{tabular}{lc}
        \hline
          Redshift source & Number of redshifts \\
        \hline
          \citet{Alabi2018} & 14\\
          \citet{SDSSDR13} (SDSS DR13) & 773\\
          \citet{Biviano1995} & 5\\
          \citet{Bravo-Alfaro2000} & 1\\
          \citet{Caldwell1993} & 1\\
          \citet{Castander2001} & 1\\
          \citet{Chiboucas2010} & 38\\
          \citet{Chiboucas2011} & 10\\
          \citet{Colless1996} & 5\\
          \citet{Cortese2003} & 2\\
          Davies et al. (1987) & 1\\
          \citet{Edwards2011} & 5\\
          \citet{Falco1999} & 2\\
          \citet{Gregory1976} & 2\\
          \citet{Gu2017} & 1\\
          \citet{Harrison2010} & 3\\
          \citet{Haynes1997} & 1\\
          \citet{Huchra2012} & 5\\
          \citet{Jangren2005}  & 1\\
          \citet{Jorgensen1995} & 1\\
          \citet{Kadowaki2017} & 2\\
          \citet{Karachentsev1990}  & 1\\
          \citet{Kourkchi2012}  & 18\\
          Lucey et al. (1991) & 2\\
          \citet{Mobasher2001} & 84\\
          \citet{Rines2001} & 3\\
          Serra et al. (in prep) & 4\\
          Smith et al. (2009) & 1\\
          This work & 59\\
          \citet{Wegner1990} & 1\\
          van Haarlem (in prep.) & 49\\
        \hline\end{tabular}
\end{table}
\onecolumn
    \begin{figure*}
    \centering
    \begin{subfigure}{0.98\textwidth}
      \includegraphics[width=\textwidth]{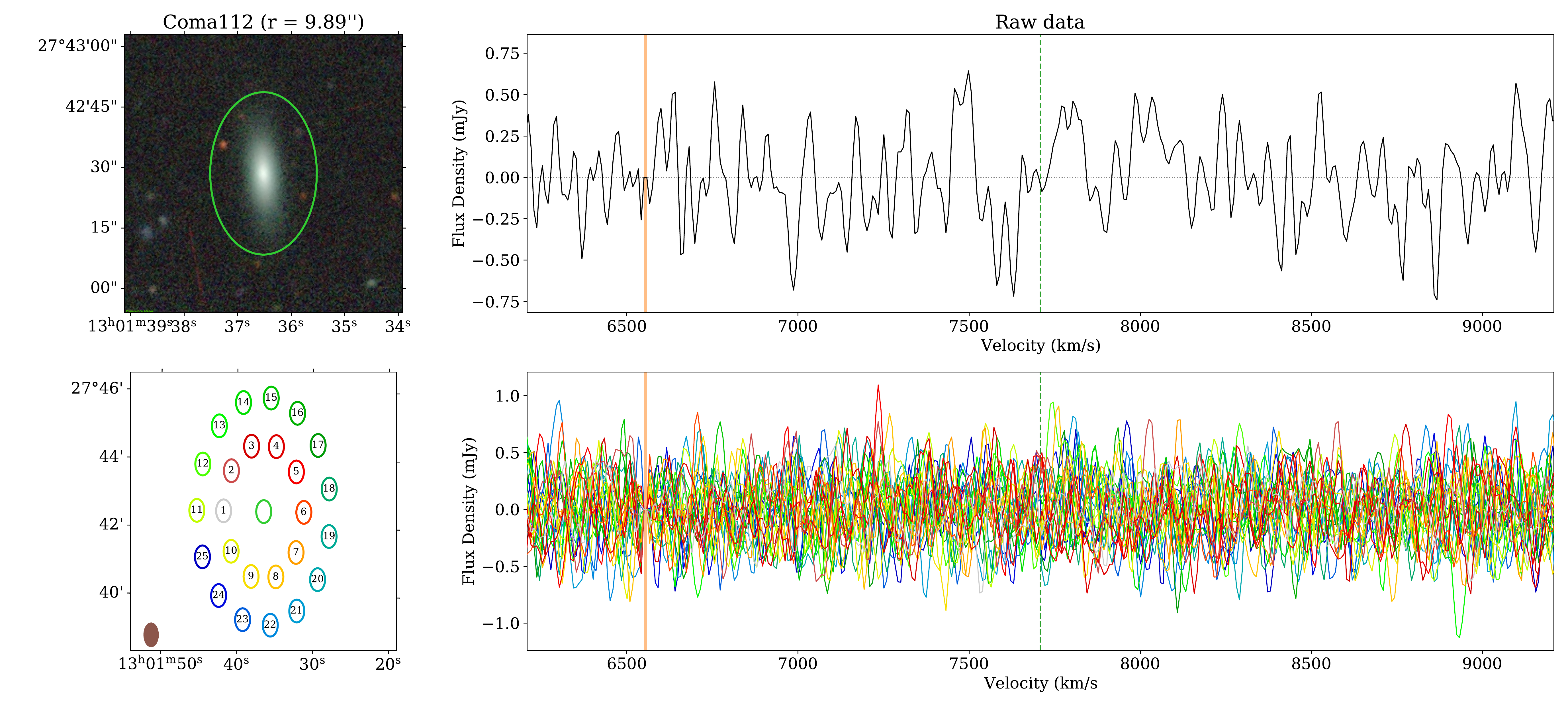}
      \caption{Source unresolved by the WRST beam}
      \label{fig:spectrumextract_unres}
    \end{subfigure}

    \begin{subfigure}{0.98\textwidth}
      \includegraphics[width=\textwidth]{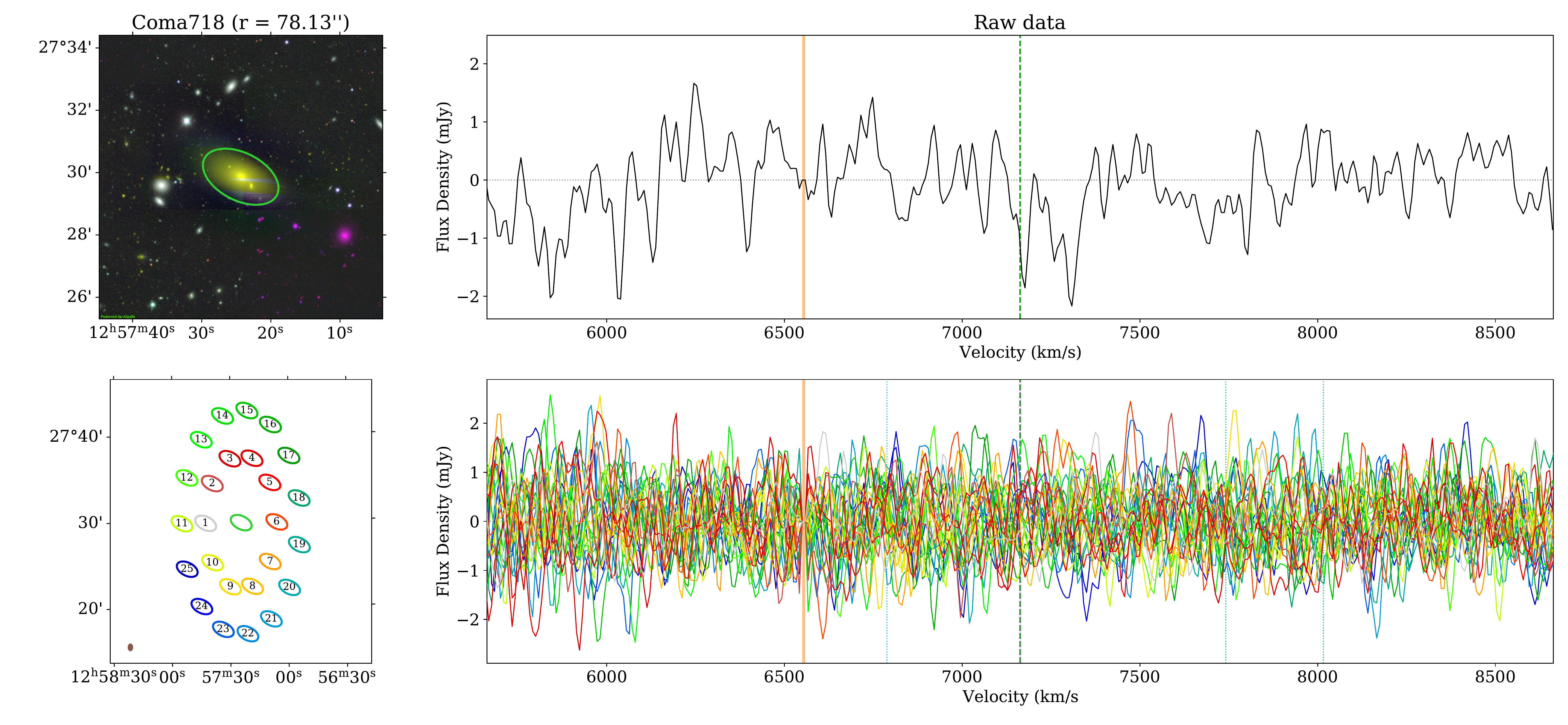}
      \caption{Source resolved by the WRST beam}
      \label{fig:spectrumextract_res}
    \end{subfigure}
    
    \caption[Spectrum extraction method]{The method used to extract the 1D \hi spectra is presented in the figures above. \autoref{fig:spectrumextract_unres} presents a galaxy which is unresolved by the WRST beam, and \autoref{fig:spectrumextract_res} a galaxy which is resolved by the beam. A colour cut-out of the galaxy with the extraction aperture overlaid in green is shown in the upper left panel. The bottom left panel shows the location of the 25 reference spectra relative to the target spectrum. The middle top panel shows the target spectrum, and the middle bottom panel shows the 25 reference spectra. In both middle panels, the vertical dashed green line indicates the redshift of the target galaxy; the orange band represents the gap between the two frequency bands.}
    \label{fig:spectra_extract}
  \end{figure*}
  
        \begin{figure*}[h]
          \centering
          \includegraphics[width=0.9\linewidth]{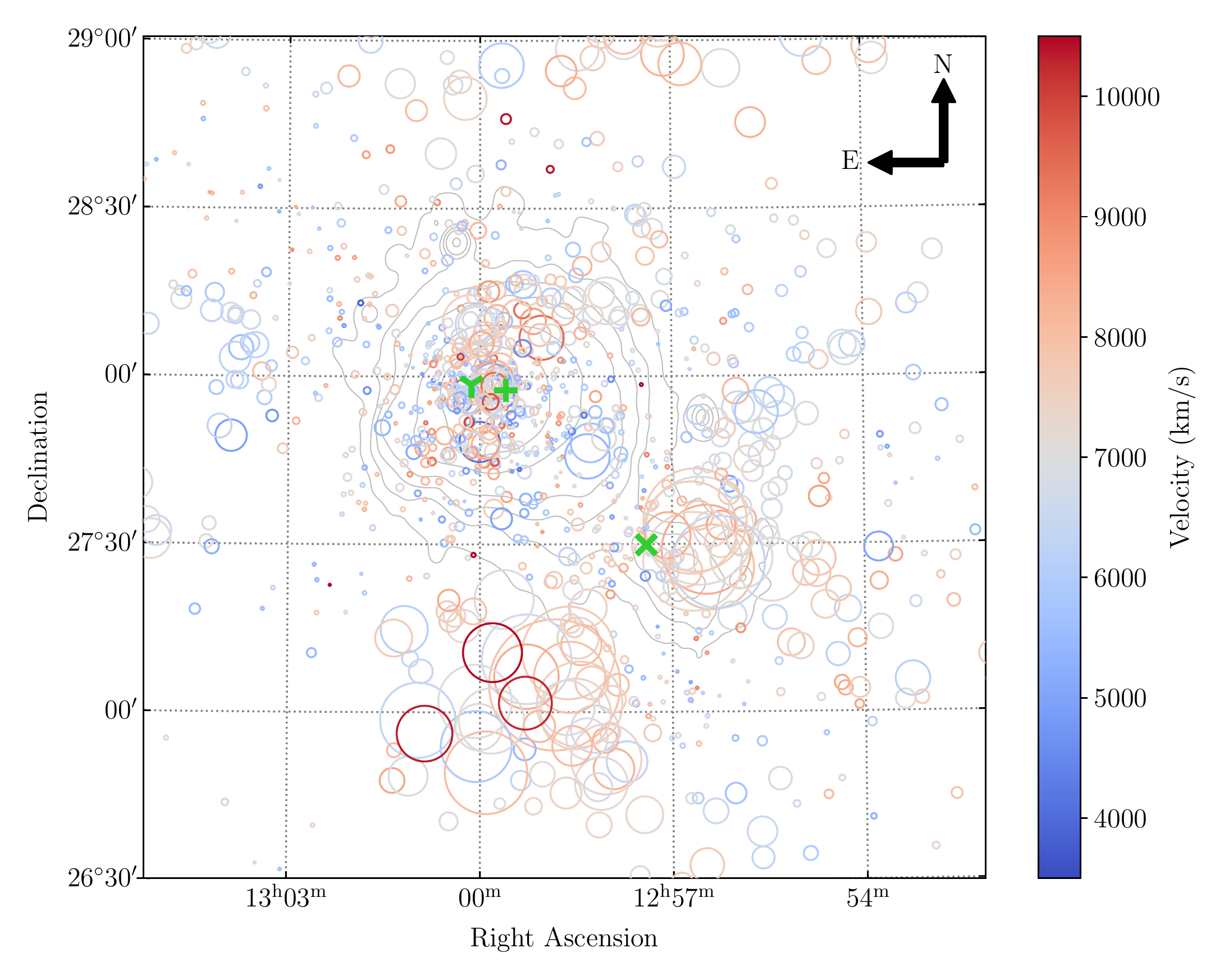}
          \caption[Dressler-Shectman Test bubble plot]{Bubble plot showing the results of the Dressler-Shectman Test for the Coma cluster. The $\delta$ values are a result of using the 25 nearest neighbours. The red/blue colours indicate the recessional velocity ($cz$) of each galaxy. The green markers represent the three cD galaxies in Coma ($+$--NGC 4874, $\curlyvee$--NGC 4889 and $\times$--NGC 4839). The background grey contours represent the x-ray emission measured by ROSAT in \numunit{0.4-2.4}{\text{keV}}.}
          \label{fig:coma_DStest}
        \end{figure*}

    \FloatBarrier
\section{Stacked Spectra}  
    \label{sec:coma_stackedspectra}

    \begin{figure*}[h]
      \centering
      \includegraphics[width=0.95\linewidth]{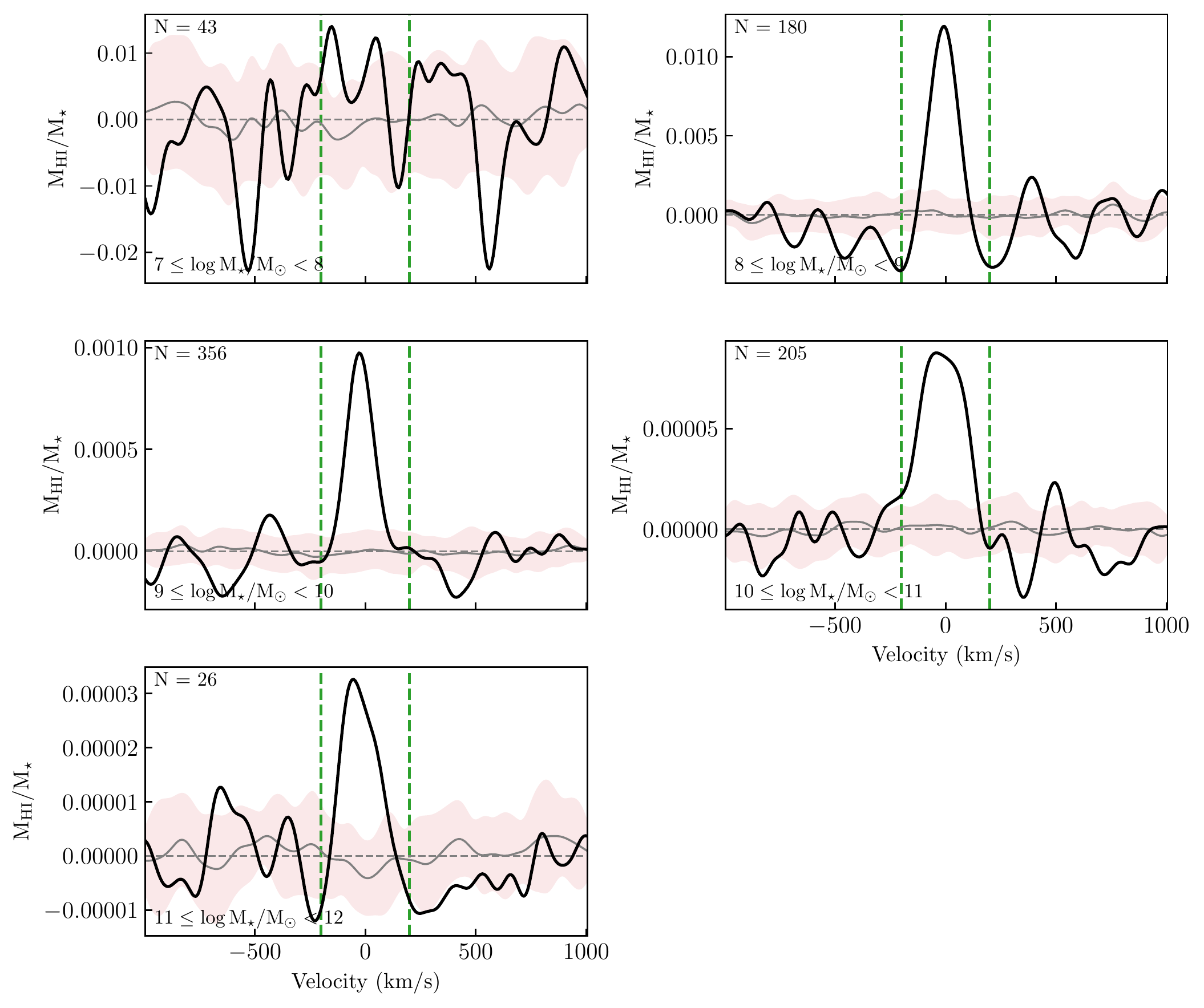}
      \caption[Stacked spectra for \autoref{fig:coma_gasfractions} -- full sample]{Stacked spectra from which the full catalogue data points in \autoref{fig:coma_gasfractions} are measured. The stacked spectra (the black line) have been smoothed to a velocity resolution of \numunit{80}{\kms}. The orange band represents the $1\sigma$ distribution from 25 stacked reference spectra. The vertical dashed green lines indicate the range over which the spectra are integrated to determine the average integrated flux density for each sample. Each panel represents a different stellar mass bin. The number of spectra in each bin are given in the top left corner of each panel.}
      \label{fig:coma_stacks_mstarfc}
    \end{figure*}

    \begin{figure*}[h]
      \centering
      \includegraphics[width=0.95\linewidth]{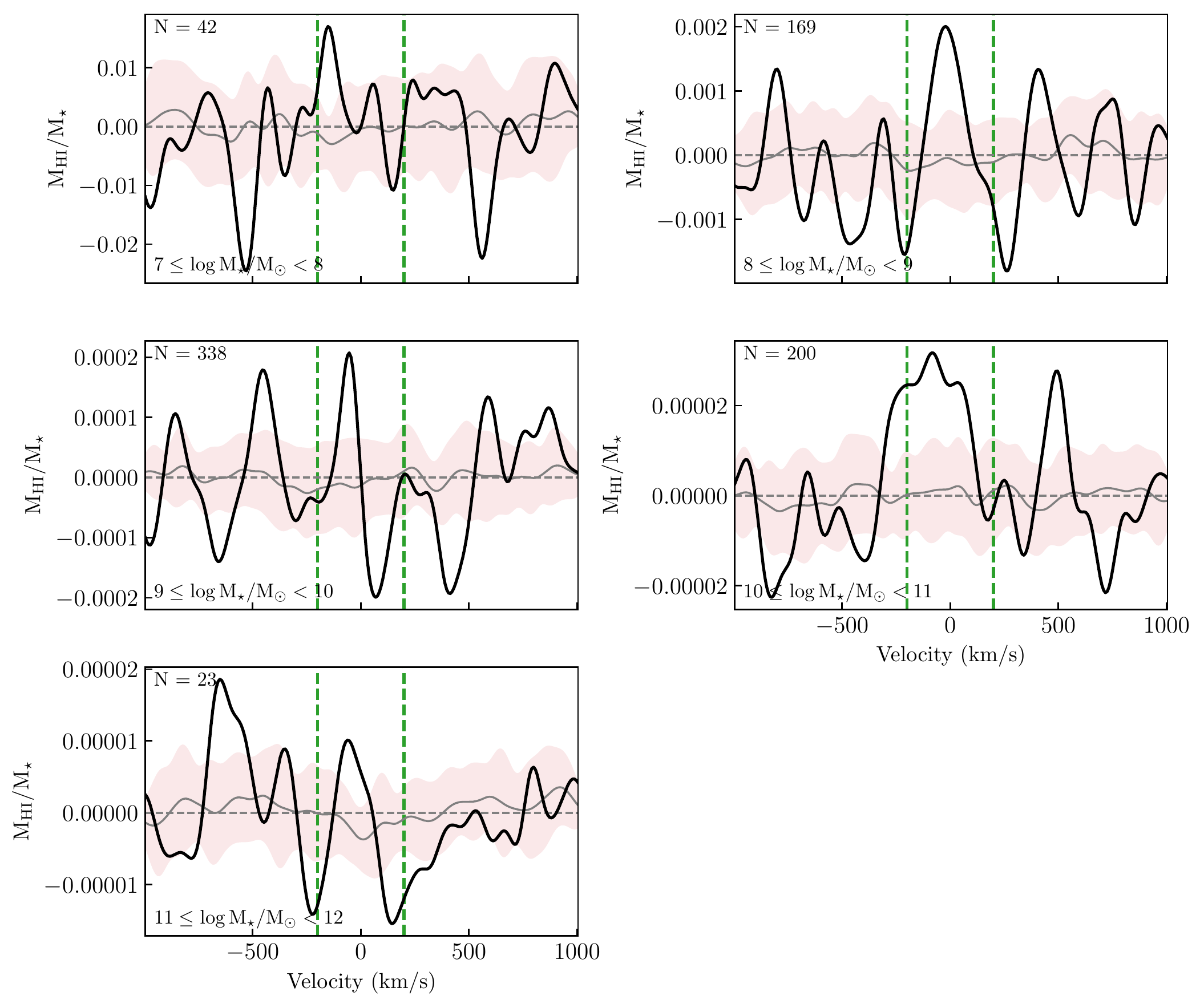}
      \caption[Stacked spectra for \autoref{fig:coma_gasfractions} -- non-detections]{Same as \autoref{fig:coma_stacks_mstarfc}, for the sample of \hi non-detections. }
      \label{fig:coma_stacks_mstarnd}
    \end{figure*}
    
    \begin{figure*}[h]
        \centering
        \includegraphics[width=\linewidth]{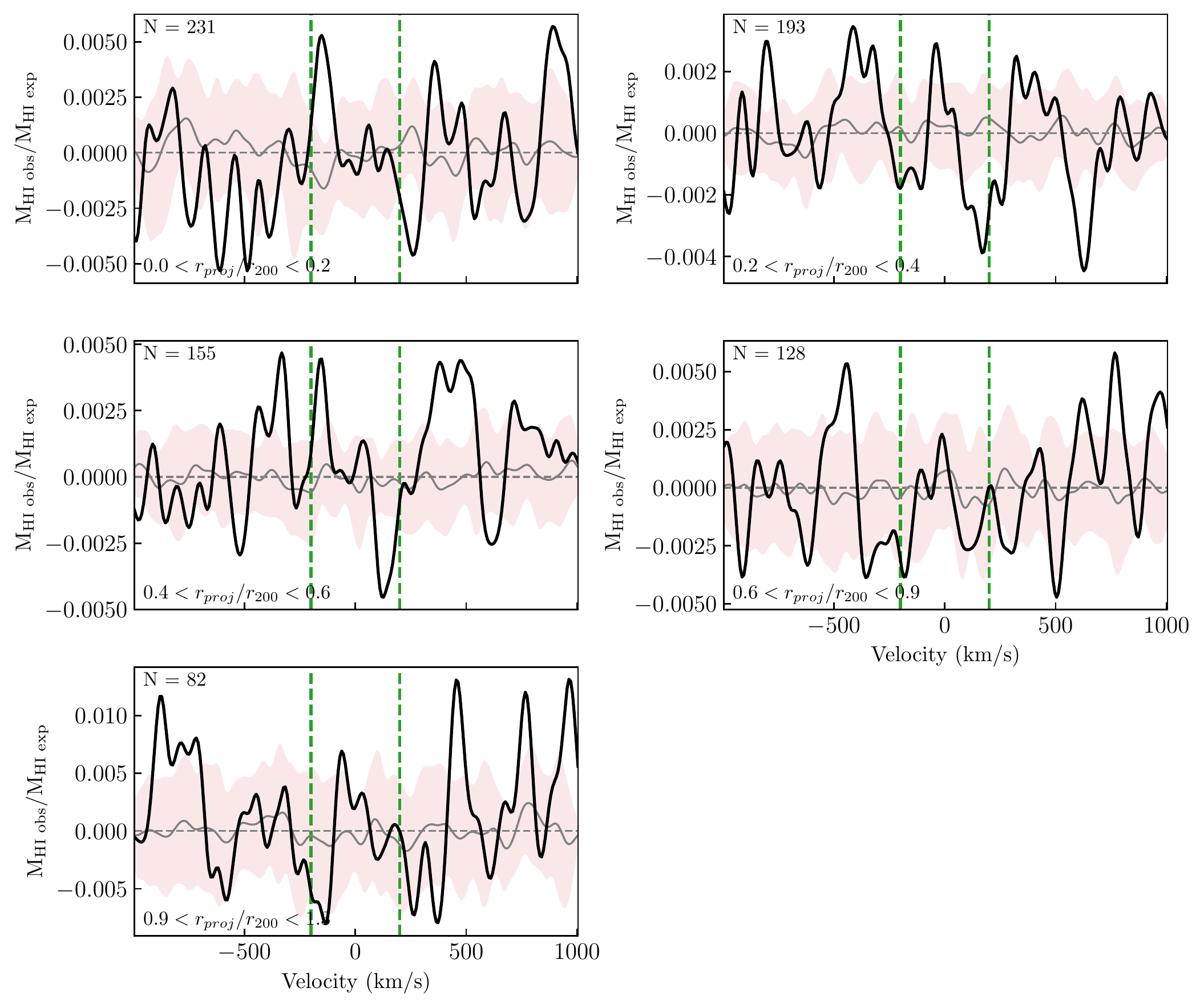}
        \caption{Coma non-detections stacked in annuli with increasing radius from the cluster centre. The stacked spectra (the black line) have been smoothed to a velocity resolution of \numunit{80}{\kms}. The orange band represents the $1\sigma$ distribution from 25 stacked reference spectra. The vertical dashed green lines indicate the range over which the spectra are integrated to determine the average integrated flux density for each sample. Each panel represents a different stellar mass bin. The number of spectra in each bin are given in the top left corner of each panel. These spectra are used to determine the average \defhi ($\defhi = \log \mathrm{M_{\hi\,\, exp}} - \log \mathrm{M_{\hi\,\, obs}}$ ) in \figref{fig:coma_defhist}.}
        \label{fig:coma_radiiDef}
    \end{figure*}
    
    \begin{figure*}[h]
        \centering
        \includegraphics[width=\linewidth]{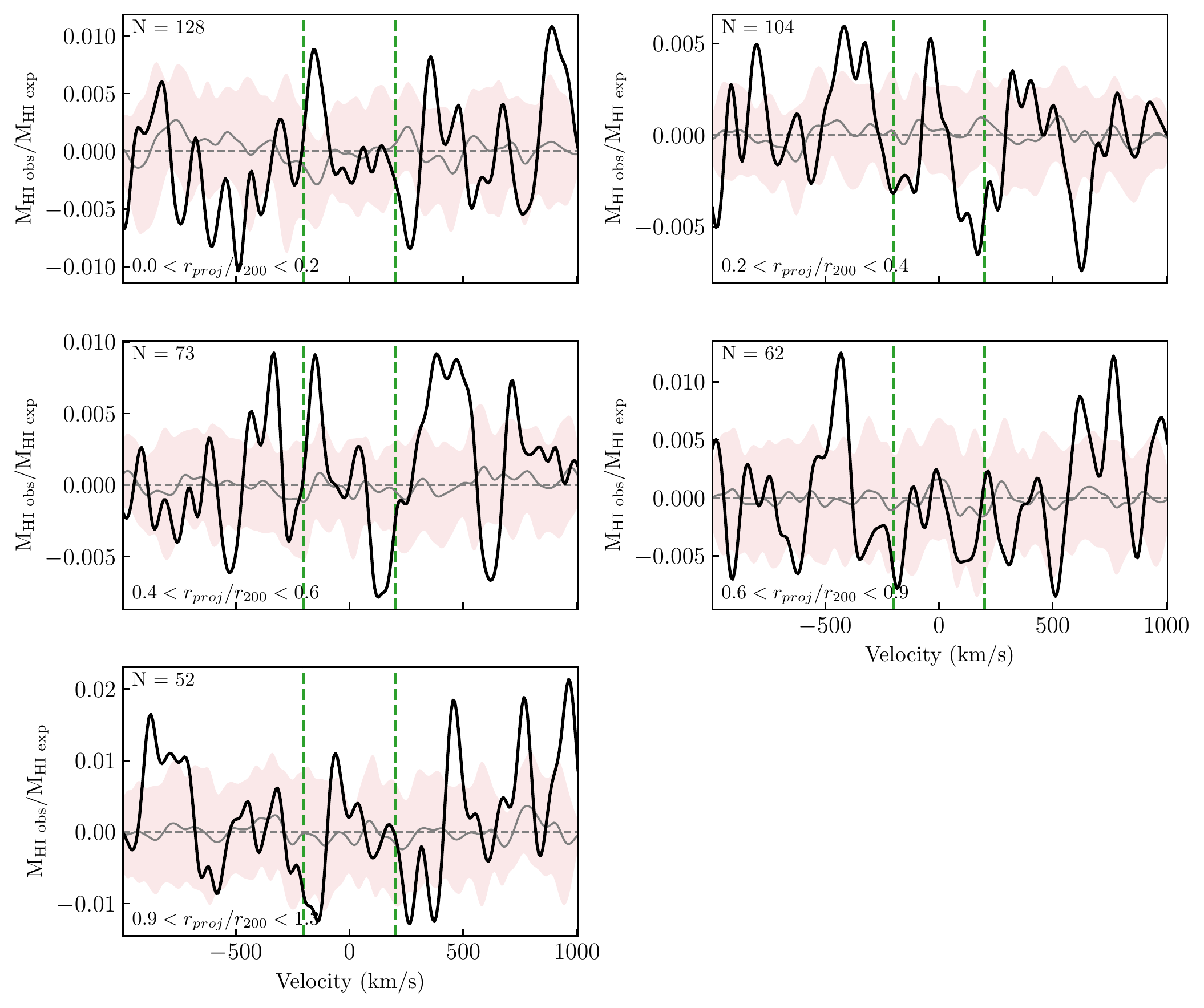}
        \caption{Same as \figref{fig:coma_radiiDef} for the sample of galaxies with $\log \mstar < 9.5$.}
        \label{fig:coma_radiiDeflsm}
    \end{figure*}
    
    \begin{figure*}[h]
        \centering
        \includegraphics[width=\linewidth]{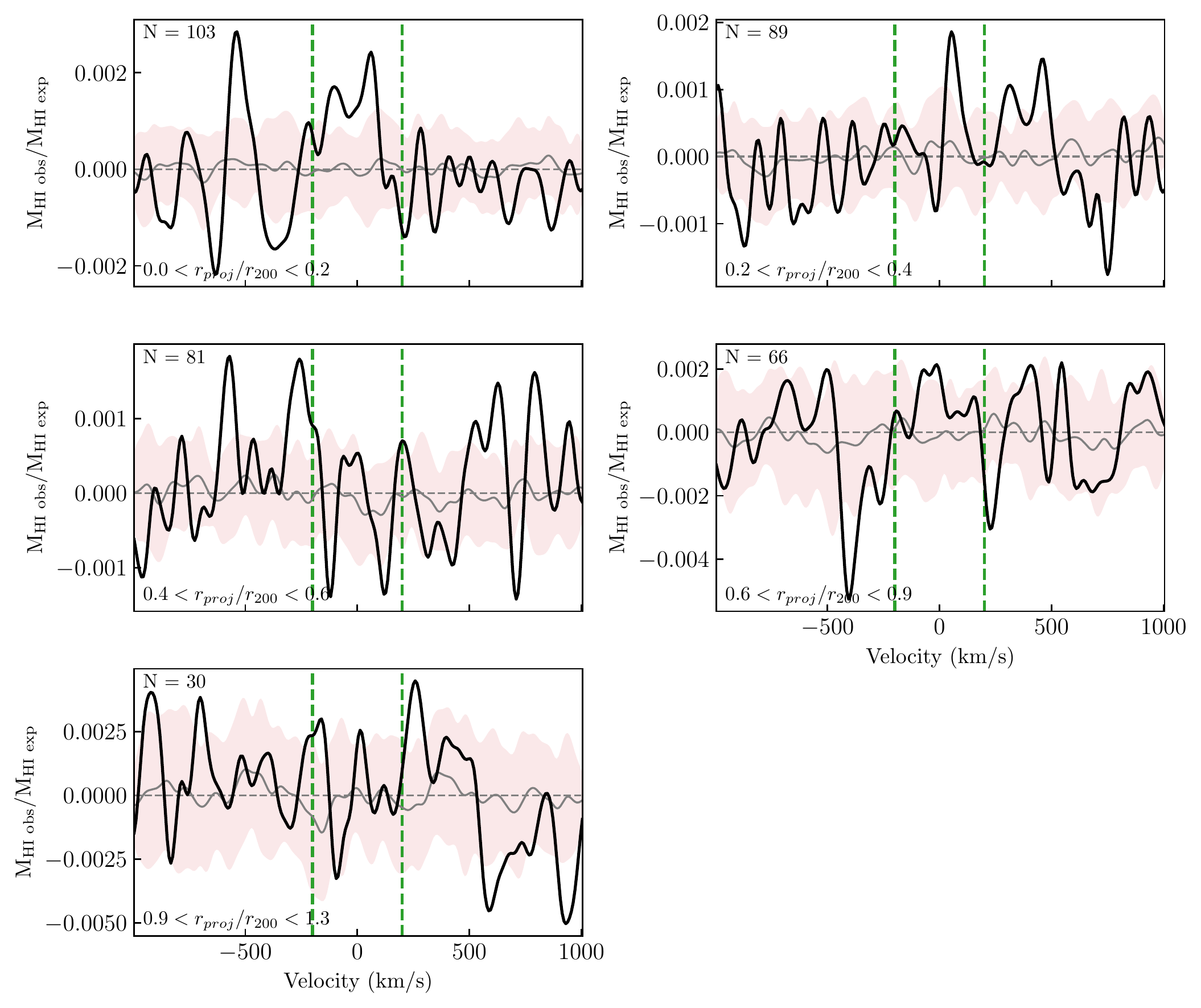}
        \caption{Same as \figref{fig:coma_radiiDef} for the sample of galaxies with $\log \mstar > 9.5$.}
        \label{fig:coma_radiiDefhsm}
    \end{figure*}
    
    \begin{figure*}[h]
        \centering
        \vspace{-12pt}
        \includegraphics[width=\linewidth, page=1]{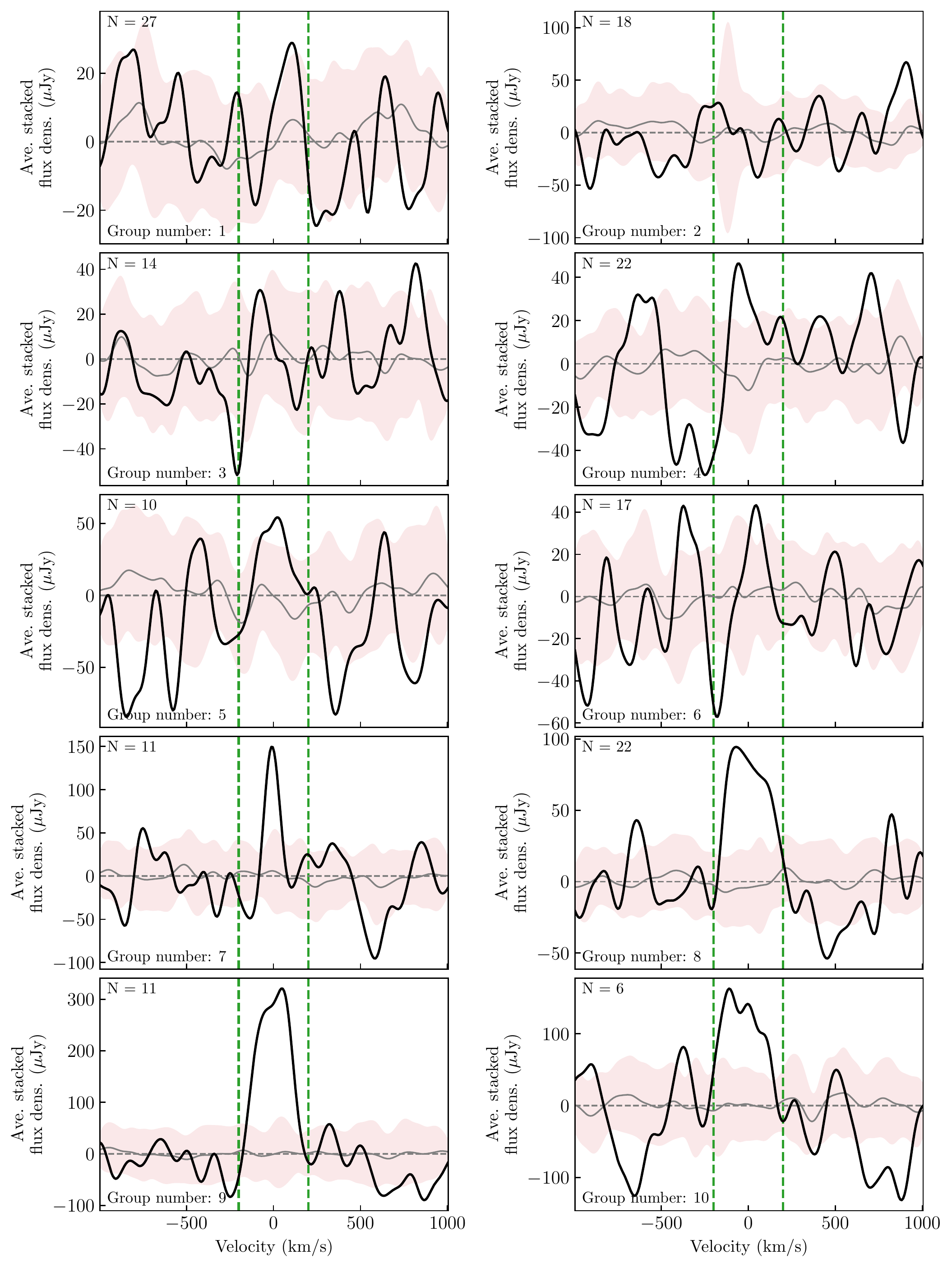}
        \vspace{-20pt}
        \caption{Stacked spectra for each of the substructures S1--S10. The stacked spectra (the black line) have been smoothed to a velocity resolution of \numunit{80}{\kms}. The orange band represents the $1\sigma$ distribution from 25 stacked reference spectra. The vertical dashed green lines indicate the range over which the spectra are integrated to determine the average integrated flux density for each sample.}
        \label{fig:coma_group_stacks1}
    \end{figure*}
    
    \begin{figure*}[h]
        \centering
        \includegraphics[width=\linewidth, page=2]{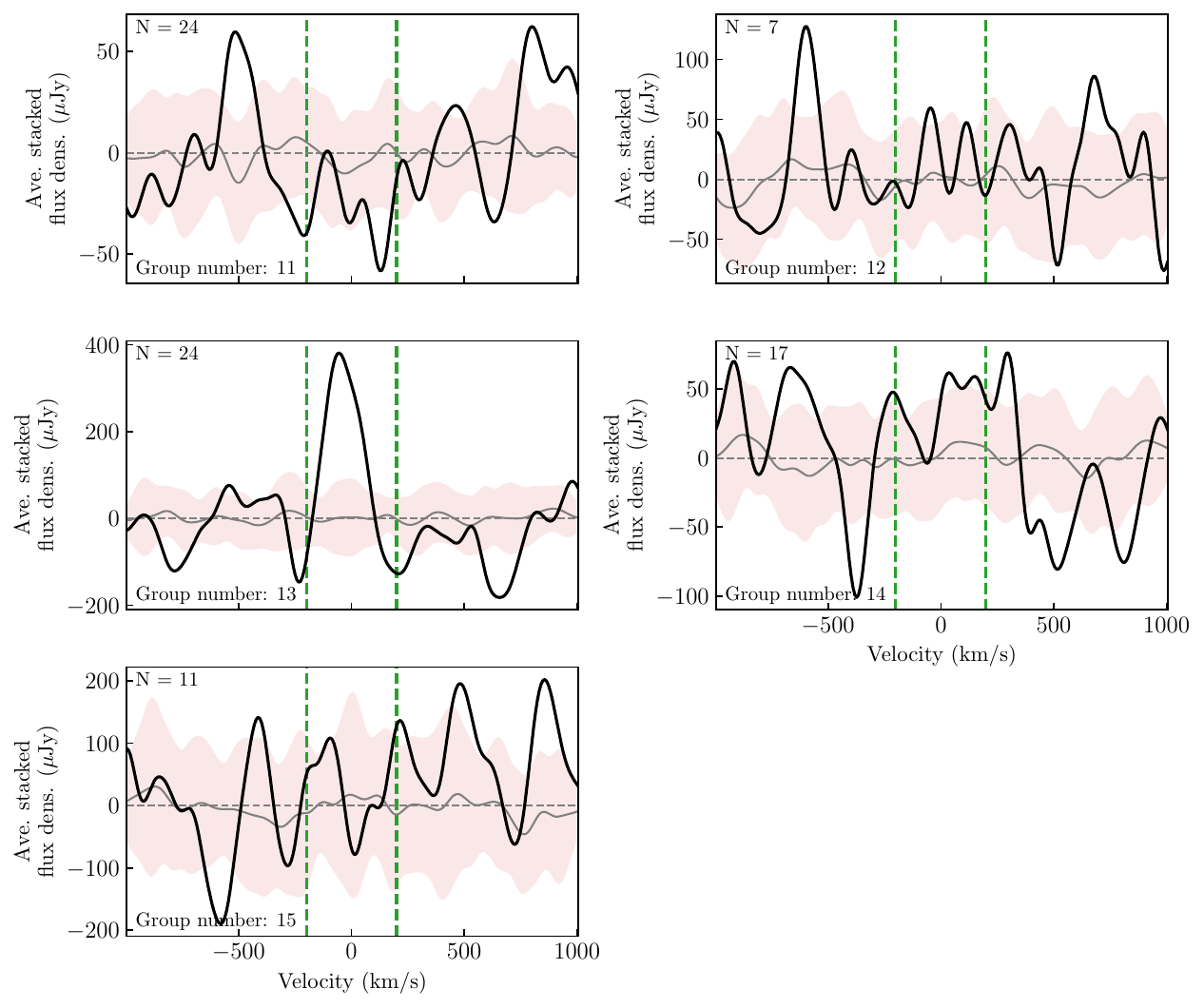}
        \caption{Same as \figref{fig:coma_group_stacks1}, but for substructures S11--S15.}
        \label{fig:coma_group_stacks2}
    \end{figure*}

\end{appendix}

\end{document}